\documentclass[sn-mathphys-num, iicol]{sn-jnl}

\usepackage{graphicx}%
\usepackage{multirow}%
\usepackage{amsmath, amssymb, amsfonts}%
\usepackage{amsthm}%
\usepackage{mathrsfs}%
\usepackage[title]{appendix}%
\usepackage{xcolor}%
\usepackage{textcomp}%
\usepackage{manyfoot}%
\usepackage{booktabs}%
\usepackage{listings}%
\usepackage{cuted}%
\usepackage{blindtext}%
\usepackage{multicol}%
\usepackage{indentfirst}%

\begin{document}

\title[Article Title]{Amplitude determination for $M M \to M M$, $M = \pi, K$ and cross-sections for $\gamma \gamma \to \pi^+ \pi^-, \pi^0 \pi^0, \pi^0 \eta$ in a chiral model}

\author[1]{\fnm{S. P.} \sur{Klevansky}}\email{spk@physik.uni-heidelberg.de}

\author[2]{\fnm{R. H.} \sur{Lemmer}}\email{deceased}

\author*[3]{\fnm{Alireza} \sur{Beygi}}\email{alirezabeygi389@gmail.com}

\affil[1]{\orgdiv{Institute for Theoretical Physics}, \orgname{Heidelberg University}, \orgaddress{\city{Heidelberg}, \postcode{69120}, \country{Germany}}}

\affil[2]{\orgdiv{Physics Department}, \orgname{University of the Witwatersrand}, \orgaddress{\city{Johannesburg}, \postcode{2050}, \country{South Africa}}}

\affil[3]{\orgdiv{Institute of Computer Science}, \orgname{Goethe University Frankfurt}, \orgaddress{\city{Frankfurt}, \postcode{60325}, \country{Germany}}}

\abstract{Dai and Pennington have performed a comprehensive analysis of essentially all pion and kaon pair production data from two-photon collisions below 1.5 GeV, including all high statistics results from Belle, as well as the older data from Mark II at SLAC, CELLO at DESY, and Crystal Ball at SLAC. Imposing the basic constraints required by analyticity, unitarity, and crossing symmetry and making use of Low's low-energy theorem for QED, they were able to extract the final-state, strong-interaction scattering amplitudes for the intermediate $\pi \pi \to \pi \pi$ and $\pi \pi \to K \overline{K}$ reactions in a model-independent fashion. In addition, they provided good fits to the respective $\gamma \gamma \to \pi \pi$ cross-sections that are known in the low-energy sector in the restricted angular range, $| \cos \theta | < 0.6 - 0.8$. Using the parameters obtained in this fashion, these authors constructed the $\gamma \gamma \to \pi \pi$ cross-sections integrated over the full angular range. In the present work, we use a version of chiral perturbation theory developed by Oller and Oset to evaluate the final-state, strong-interaction amplitudes theoretically, and we compare our low-energy QCD-based results with the amplitudes extracted by Dai and Pennington. We also calculate the $\gamma \gamma \to \pi \pi$ cross-sections (integrated over the full angular range) and compare them with those obtained by Dai and Pennington. These calculations give a more detailed insight into the fit of chiral perturbation theory, not just to the measured $\gamma \gamma \to \pi \pi$ cross-sections, as is usually presented, but rather to a higher level of detail through the available analysis of the experimental data for the underlying final-state, strong-interaction, meson--meson scattering amplitudes $\pi \pi \to \pi \pi$ and $\pi \pi \to K \overline{K}$ themselves. The fits appear to be sensible over the energy range considered. The detailed calculations of strong-interaction transition matrices, as presented in this paper, also pave the way to address the possible presence of the postulated kaonium atom $K^+ K^-$ in the cross-section.}


\maketitle

\section{Introduction}
\label{sec1}

Photon--photon to meson--meson cross-sections have been measured by several experimental groups over the last decades \cite{Be07, Be08, Be09, Be13, HM90, TOest90, Hyams, Cohen80, Etkin82, Batley10}. The high statistics experimental data obtained by the Belle Collaboration at KEKB for $\gamma \gamma \to \pi^+ \pi^-$ \cite{Be07}, $\gamma \gamma \to \pi^0 \pi^0$ \cite{Be08}, and $\gamma \gamma \to \pi^0 \eta$ \cite{Be09} cross-sections, plus the similar high-quality data for $\gamma \gamma \to K^+ K^-$ and $\gamma \gamma \to K^0 \overline{K}^0$ for $\gamma \gamma$ center-of-mass collision energies up to $\sim 1$ GeV \cite{Cohen80, Etkin82, Batley10} have given new impetus to the field and can provide important new information with which to probe the possible quark structure of the light isoscalar $f_0 ( 500 )$, $f_0 ( 980 )$, and isovector $a_0 ( 980 )$ scalar mesons \cite{Pelaez, NNA12, MRP08, MH2011, AT2017, JRP2021}. Dai and Pennington have performed a comprehensive {\it amplitude analysis} of the processes $\gamma \gamma \to \pi^+ \pi^-$, $\pi^0 \pi^0$, and $K \overline{K}$ below 1.5 GeV \cite{DaiPenn14}. Using all available experimental data, they have extracted the associated final-state, strong-interaction transition matrices, $\pi \pi \to \pi^+ \pi^-, \pi^0 \pi^0, K \overline{K}$ in a model-independent fashion, using only properties of analyticity, unitarity, and crossing symmetry and Low's low energy theorem for QED. Their fits pertain to the experimental data that are measured over a restricted angular range, $| \cos \theta | < 0.6 - 0.8$. Having determined all parameters, they are able to construct the cross-sections for $\gamma \gamma \to \pi^+ \pi^-, \pi^0 \pi^0$ that would be expected after integrating over the full angular range.

Such developments open up several intriguing possibilities from a theoretical point of view. (a) First, the precise knowledge of the final-state, strong-interaction transition matrices can be used to test the predictions of low-energy QCD, or at least differentiate between various models thereof. (b) In the energy range studied, it opens up the possibility of examining the viability of detailed combined structures that potentially can form. Various strong-interaction models with different structural properties have already been explored in some detail for the light isoscalar and isovector mesons. These include, for example, descriptions with simple $q \bar q$ pairs \cite{LM00, NATR96, VD96, RDS98, MDS04}, more complex $q^2 \bar q^2$ states \cite{NNA89}, or a $K \overline{K}$ molecular structure \cite{WIsgur82, TB85, SU390, JAO03, KLS04, TBR08, RHL09} for the $f_0 ( 980 )$ and $a_0 ( 980 )$ in particular. Now, in addition to the strong interaction, which is known through the analysis of \cite{DaiPenn14}, electromagnetic effects can be incorporated, and, for example, the resonance formed in the production and subsequent decay of the $K^+ K^-$ hadronic atom kaonium \cite{AMG93, BKER95, SVBKER96, SPKR11} can be studied theoretically. This, in turn, may become accessible experimentally. As a reminder, note that attractive Coulomb interactions are crucial for the formation of kaonium, which, in turn, implies isospin breaking. However, because of the disparate length scales over which the strong and Coulomb interactions operate, isospin breaking is confined to the region outside the strong interaction range. By assuming this physical assumption and using meson--meson interaction amplitudes, the authors of \cite{SPKR11} have been able to investigate strong interaction effects on the binding (and decay) of kaonium, and have found lifetimes of $(2.2 \pm 0.9) \times 10^{-18} \; $s$ $ for its ground state. More recently, the existence of the 2p state of kaonium has been proposed in \cite{PL2020}, by analyzing the data obtained by the CMD-3 experiment on the $e^+ e^- \to K^+ K^-$ process.

The present paper aims to examine (a) above and to determine how well leading-order chiral perturbation theory (ChPT) \cite{GL84, GE95}, taken together with QED to calculate the final-state, strong-interaction and electromagnetic transition matrices, serves to give a good description of the final-state, strong-interaction transition matrices as compared with the model-independent curves extracted by Dai and Pennington from experiment \cite{DaiPenn14}. Our calculated final-state, strong-interaction transition matrices turn out both qualitatively and quantitatively to be in reasonable agreement for both the individual real and imaginary parts. We also calculate the full cross-sections for the $\gamma \gamma \to \pi^+ \pi^-$ and $\pi^0 \pi^0$ reactions, incorporating the electromagnetic contributions, and again find a reasonable, but not perfect, agreement with the extracted curves of Dai and Pennington. The question of new structures (b), such as the presence of kaonium appearing in the cross-section, which requires a solid knowledge of the strong-interaction transition matrices, as presented here, is left as a subject for our future paper. In a nutshell, we will search for the kaonium as a sharp resonance possibly accompanying the $f_0(980)$ in the processes $\gamma \gamma \to \pi^0 \pi^0$ and $\gamma \gamma \to \pi^0 \eta$. This will require the modifications of the cross-sections of these processes, which are presented in the current paper, to include the formation of kaonium, that essentially boils down to the inclusion of isospin breaking in the transition amplitudes. In the corresponding calculated cross-sections, one would expect the same behavior as that of, for example, $\gamma \gamma \to \pi^0 \eta$, except in the close vicinity of kaonium resonances, where the isospin breaking is significant.

For our aims in the present paper, both for evaluating the full transition matrices and for calculating the photon--photon to meson pair cross-sections, we require a detailed analysis of the underlying electromagnetic interaction as well as the strong-interaction component through meson--meson scattering processes. Such studies are not uncommon: The first calculation of two-pion production in photon--photon collisions via ChPT presented against data was performed by Bijnens and Cornet \cite{BiCo88} when the first data from the Crystal Ball experiment were made available. Thereafter, in a particularly clear fashion, Oller and Oset \cite{JAO97} extracted meson--meson interactions within the pseudoscalar meson $SU ( 3 )$ flavor octet from the Lagrangian given by leading order ChPT \cite{GL84, GE95} as the appropriate theoretical realization of low energy QCD. They then used these interactions as input for the Lippmann--Schwinger equation to provide a non-perturbative calculation of the pseudoscalar meson--meson scattering and reaction amplitudes.

It is, however, important to bear in mind that the validity of the leading order ChPT results is restricted to center-of-mass collision energies up to $\sim {\cal O} ( 1 \; $GeV$ )$. A glance at the two-photon collision data \cite{Be07, Be08, Be09} shows that, while the $f_0 ( 500 )$, $f_0 ( 980 )$, and $a_0 ( 980 )$ again appear quite naturally in the ChPT calculations as dynamically generated resonances \cite{JAOgg, JAO20081, JAO20082} below 1 GeV center-of-mass total energy, with energies and widths compatible with the experiment in the total cross-section of the relevant reaction channels, the dominance of the wide $f_2 ( 1270 )$ and $a_2 ( 1320 )$ resonances eventually overshadow the ChPT contribution at higher energies. Whilst not important for studying the $\gamma \gamma \to$ kaonium production process, we remark that when the ChPT transition amplitudes are supplemented by contributions from the above two resonances in the parametrized form \cite{Albrecht89}, both being interpreted as $d$-wave, helicity $\lambda = 2$ states, plus the exchange of vector and axial vector octet resonances in the $u$ and $t$ channels \cite{DH93}, there is good agreement with the available photon--photon collision data over the entire energy range from the two-meson threshold to $\sim 1.4 \; $MeV$ $ in the center-of-mass system.

A decade after the first ChPT comparison with data \cite{BiCo88}, Oller and Oset recalculated the scattering cross-sections in ChPT, including the $f_2$ and $a_2$ mesons, and compared these with the then available data \cite{JAOgg}. In our comparison with the precision data available, we follow their approach.

This article is arranged as follows. Section~\ref{sec2} addresses the calculation of the $T$-matrices for the photon--photon interactions. In Secs.~\ref{sec2a}, \ref{sec2b}, and \ref{sec2c}, we build up the Born contributions, the contributions containing meson--meson scattering through ChPT, and the resonant and axial contributions that cannot be described by ChPT, respectively. These are collated in Sec.~\ref{sec2d}. In Sec.~\ref{sec3}, we compare our calculated amplitudes and cross-sections with the extracted fits of \cite{DaiPenn14} and experimental results. We summarize and conclude in Sec.~\ref{sec4}.

\section{Production amplitudes for the processes $\gamma \gamma \to \pi^+ \pi^-$, $\pi^0 \pi^0$, $\pi^0 \eta$, and $K^+ K^-$}
\label{sec2}

\begin{figure*}

\centering

\includegraphics[scale = 1]{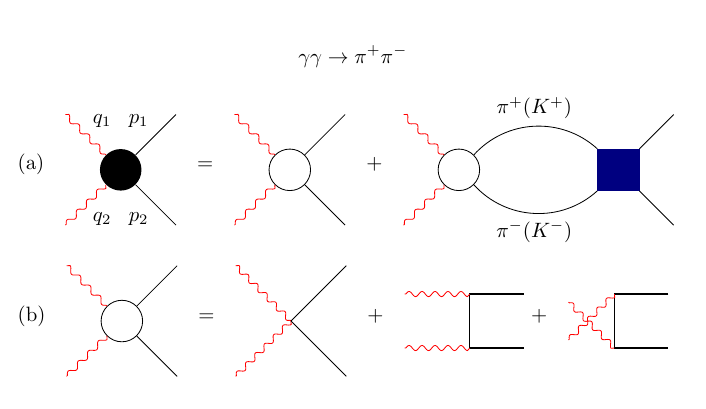}

\caption{Scattering amplitudes for $\gamma ( q_1 ) + \gamma ( q_2 ) \to m_1 ( p_1 ) + m_2 ( p_2 )$ collisions producing meson pairs of masses $( m_1, m_2 )$ with incoming and outgoing momenta $( q_1, q_2 )$ and $( p_1, p_2 )$, respectively. (a) The full amplitude (filled circle). This amplitude includes strong interactions in the final state as given by the 4-point, meson--meson scattering $T$-matrix (filled box diagram). (b) The Born term (open circle) only involves the electromagnetic coupling vertices of photons to charged mesons and is only present for $\pi^+ \pi^-$ or $K^+ K^-$ in the final state.}

\label{fF1}
\end{figure*}

The theoretical basis for evaluating the $\gamma \gamma \to m_1 m_2$ processes has been studied in different contexts or models before, see for example \cite{JAOgg, DaiPenn14} and references cited therein, and involves both the electromagnetic coupling of the photons to (charged) mesons that is determined through QED, as well as QCD for the final-state strong interactions between the mesons themselves. As the latter cannot be extracted directly from QCD itself, we use ChPT as an appropriate realization of the strong interactions in the low-energy sector. These processes are represented graphically in Fig.~\ref{fF1} by the Feynman diagrams for the $\gamma \gamma \to m_1 m_2$ transition amplitude for incoming photon four-momenta and helicities $( q_1, \lambda_1 )$ and $( q_2, \lambda_2 )$ leading to outgoing mesons with four-momenta $( p_1, p_2 )$. The filled circle diagram in panel (a) of Fig.~\ref{fF1} denotes the full transition amplitude tensor $i T^{\mu \nu}_{\gamma \gamma \to m_1 m_2}$. This itself is resolved into two parts: a direct coupling of the electromagnetic interaction to the mesons plus a term in which both the electromagnetic and strong interactions play a role. The first is the Born term $i T^{\mu \nu}_{B ; \gamma \gamma \to m_1 m_2}$, denoted in the figure as an open circle, and is only present when photons couple to charged meson pairs $m_+ m_-$ in the final state. The second diagram, denoted as $i T^{\mu \nu}_{S ; \gamma \gamma \to m_1 m_2}$, includes the contribution from the strong meson--meson interactions in the final state. The direct coupling of the electromagnetic interaction to the mesons via the Born term is again broken down into three individual terms, also shown in this figure, see panel (b), and which contain one-meson exchange.

As it stands, low-energy ChPT does not account sufficiently for the $f_2 ( 1270 )$ and $a_2 ( 1320 )$ resonances, which have large widths that extend well into the region below 1 GeV. These are accommodated in our formalism via parametrization. We thus proceed as follows: in Sec.~\ref{sec2a}, we start with the Born term to set our notation. We then evaluate contributions from the final-state strong interactions using ChPT in Sec.~\ref{sec2b}. In Sec.~\ref{sec2c}, we give the parametrization for resonant, non-ChPT terms. Then, in Sec.~\ref{sec2d}, we build up the full $T$-matrix, that consists of the amplitudes

\begin{strip}
\rule{\dimexpr(0.5\textwidth-0.5\columnsep-0.4pt)}{0.4pt}%
\begin{equation}
T^{\mu \nu}_{\gamma \gamma \to m_1 m_2} = T^{\mu \nu}_{B; \gamma \gamma \to m_1 m_2} + T^{\mu \nu}_{S; \gamma \gamma \to m_1 m_2} + T^{\mu \nu}_{R ( A ); \gamma \gamma \to m_1 m_2},
\label{e:tmunu}
\end{equation}
\par
\hfill
\rule[0.5\baselineskip]{\dimexpr(0.5\textwidth-0.5\columnsep-1pt)}{0.4pt}

\begin{multicols}{2}
with $\mu, \nu = 0, \dots, 3$, and specify the cross-sections. Throughout this work, we use natural units, where $\hbar = c = 1$ and the charge $e^2 / 4 \pi = \alpha$.

\subsection{Electromagnetic contributions to $\gamma \gamma \to m^+ m^-$ in the helicity basis}
\label{sec2a}

\hspace{\parindent} The process $\gamma \gamma \to \pi^+ \pi^-$ due to electromagnetic interactions (Born approximation) has been well-studied in the literature, see for example \cite{DH93}. In this subsection, we provide important results, giving sufficient detail to make this paper self-contained, allowing the reader to follow the calculations. These are directly applicable to the process $\gamma \gamma \to K^+ K^-$. In general, the transition amplitude in the Born approximation corresponding to the second diagram in panel (a) of Fig.~\ref{fF1}, or all diagrams in panel (b) of Fig.~\ref{fF1} leads to the expression
\end{multicols}

\rule{\dimexpr(0.5\textwidth-0.5\columnsep-0.4pt)}{0.4pt}%
\begin{equation}
T^{\mu \nu}_{B; \gamma \gamma \to m^+ m^-} = e^2 \left [ 2 g^{\mu \nu} + \frac{( 2 p_1 - q_1 )^\mu ( 2 p_2 - q_2 )^\nu}{( p_1 - q_1 )^2 - m_\pm^2} + \frac{( 2 p_1 - q_2 )^\nu ( 2 p_2 - q_1 )^\mu}{( p_1 - q_2 )^2 - m_\pm^2} \right ],
\label{eq2}
\end{equation}

\begin{multicols}{2}
where $m_\pm$ are the (common) masses of the final-state mesons. This expression has to be contracted with polarization vectors $\epsilon_\mu$ and $\epsilon_\nu$. For an explicit evaluation, without loss of generality, we use a standard choice \cite{FHM84} $\epsilon_0 = 0$ and the 3-vectors of helicity $\lambda_1$, $\lambda_2$, ${\bf e}_{\lambda_1} ( 1 )$, ${\bf e}_{\lambda_2} ( 2 )$, both oriented along right-handed orthogonal axes $x y$ perpendicular to the photon momentum vector in the $z$ direction, ${\bf e}_{\lambda_1} ( 1 ) = {\bf \hat{i}}, {\bf e}_{\lambda_2} ( 2 ) = {\bf \hat{j}}$, and ${\bf e}_{\lambda_1} ( 1 ) \cdot {\bf e}_{\lambda_2} ( 2 ) = \delta_{{\lambda_1}, {\lambda_2}}$ to fulfill the Lorentz condition. This leads to the expression
\end{multicols}
\rule{\dimexpr(0.5\textwidth-0.5\columnsep-0.4pt)}{0.4pt}%
\begin{align}
( e_{\lambda_1} ( 1 ) )_i T^{i j}_{B; \gamma \gamma \to m^+ m^-} ( e_{\lambda_2} ( 2 ) )_j = -2 e^2 \biggl [ &{\bf e}_{\lambda_1} ( 1 ) \cdot {\bf e}_{\lambda_2} ( 2 ) \nonumber \\
&+ \frac{( {\bf p}_1 \cdot {\bf e}_{\lambda_1} ( 1 ) ) ( {\bf p}_2 \cdot {\bf e}_{\lambda_2} ( 2 ) )}{p_1 \cdot q_1} + \frac{( {\bf p}_2 \cdot {\bf e}_{\lambda_1} ( 1 ) )( {\bf p}_1 \cdot {\bf e}_{\lambda_2} ( 2 ) )}{p_1 \cdot q_2} \biggr ],
\label{e: 3polcon}
\end{align}
\par
\hfill
\rule[0.5\baselineskip]{\dimexpr(0.5\textwidth-0.5\columnsep-1pt)}{0.4pt}
\begin{multicols}{2}
where the indices $i$, $j$ can take on the values 1, 2, 3.

\hspace{\parindent} In the center-of-mass system, the incoming photon and outgoing meson lines in panel (b) of Fig.~\ref{fF1} have four-momenta $q_1 = ( P_0 / 2, {\bf q} )$, $q_2 = ( P_0 / 2, - {\bf q} )$, $p_1 = ( P_0 / 2, {\bf p} )$, and $p_2 = ( P_0 / 2, - {\bf p} )$, where $\sqrt s = P_0$ is the total collision energy. Using this, the contracted Born amplitude becomes
\end{multicols}
\rule{\dimexpr(0.5\textwidth-0.5\columnsep-0.4pt)}{0.4pt}%
\begin{align}
\left ( T_{B; \gamma \gamma \to m^+ m^-} \right)_{\lambda_2 \lambda_1} &= ( e^*_{\lambda_2} ( 2 ) )_i T^{i j}_{B; \gamma \gamma \to m^+ m^-} ( e_{\lambda_1} ( 1 ) )_j \nonumber \\
&= -2 e^2 \left [ {\bf e}^*_{\lambda_2} ( 2 ) \cdot {\bf e}_{\lambda_1} ( 1 ) - 2 \frac{( {\bf v} \cdot {\bf e}^*_{\lambda_2} ( 2 ) ) ( {\bf v} \cdot {\bf e}_{\lambda_1} ( 1 ) )}{1 - v^2 \cos^2 \theta} \right ],
\label{e:T3cms}
\end{align}
\par
\hfill
\rule[0.5\baselineskip]{\dimexpr(0.5\textwidth-0.5\columnsep-1pt)}{0.4pt}
\begin{multicols}{2}
where ${\bf e}_{\lambda_1} ( 1 )$ is associated with particle 1 with incoming momentum ${\bf q}$ and ${\bf e}_{\lambda_2} ( 2 )$, with particle 2 with incoming momentum $- {\bf q}$. Also, the center-of-mass velocity is $v = 2 p / P_0 = \sqrt{1 - 4 m_{\pm}^2 / s}$, and $\cos \theta = {\bf p} \cdot {\bf q} / p q$ gives the polar angle of the scattering direction of the outgoing meson ${\bf p}$ relative to the incoming photon ${\bf q}$. We choose to evaluate (\ref{e:T3cms}) for the Born amplitudes in the chiral helicity basis that is defined by ${\bf e}_{R, L} ( l ) = \mp ( {\bf e}_1 ( l ) \pm i {\bf e}_2 ( l ) ) / \sqrt{2}$, $l = 1, 2$. Note that in the center-of-mass system, the total helicity of the colliding photon pair can only take on the values $\lambda = 0$ or 2, and this label is sufficient to characterize the contracted $T$-matrices, which we denote as $T^{( \lambda )}$. Then the individual contracted amplitudes are easily found to be \cite{JAOgg},
\end{multicols}
\rule{\dimexpr(0.5\textwidth-0.5\columnsep-0.4pt)}{0.4pt}%
\begin{equation}
\left ( T^{( \lambda = 0 )}_{B; \gamma \gamma \to m^+ m^-} \right )_{R ( 2 ) R ( 1 )} = \left ( T^{( \lambda = 0 )}_{B; \gamma \gamma \to m^+ m^-} \right )^*_{L ( 2 ) L ( 1 )} = -2 i e^2 \frac{1 - v^2}{1 - v^2 \cos^2 \theta},
\label{e:bornbasis1}
\end{equation}
\begin{equation}
\left ( T^{( \lambda = 2 )}_{B; \gamma \gamma \to m^+ m^-} \right )_{L ( 2 ) R ( 1 )} = \left ( T^{( \lambda = -2 )}_{B; \gamma \gamma \to m^+ m^-} \right )^*_{R ( 2 ) L ( 1 )} = 2 i e^2 \frac{v^2 \sin^2 \theta e^{2 i \phi}}{1 - v^2 \cos^2 \theta}.
\label{e:bornbasis2}
\end{equation}

\begin{multicols}{2}
\hspace{\parindent} An expansion of the $T^{( \lambda )}_{B; \gamma \gamma \to m^+ m^-}$ in spherical harmonics $Y_{J, \lambda} (\theta, \phi)$ for each total helicity $\lambda$ yields the partial contracted amplitudes $T^{( J, \lambda )}_{B; \gamma \gamma \to m^+ m^-}$, which can be used to identify the leading $s$- and $d$-wave contributions to (\ref{e:bornbasis1}) and (\ref{e:bornbasis2}). To this end, we write (\ref{e:bornbasis1}) as
\begin{equation}
T^{( 0 )}_{B; \gamma \gamma \to m^+ m^-} = \sum_{J = 0, 2, 4, \cdots} T^{( J, 0 )}_{B; \gamma \gamma \to m^+ m^-}.
\label{e:helicity0}
\end{equation}
The $( J, \lambda )$ partial-wave amplitude for helicity zero is identified as
\begin{align}
&T^{( J, 0 )}_{B; \gamma \gamma \to m^+ m^-} = -2 i e^2 ( 1 - v^2 ) / v \nonumber \\
&\times \sqrt{4 \pi} ( 2 J + 1 )^{1 / 2} Q_J ( 1 / v ) Y_{J, 0} ( \theta, \phi ),
\end{align}
where $1 / v > 1$ and $Q_J ( 1 / v )$ is a Legendre function of the second kind \cite{AS}. We have also dropped the chiral indices, which are no longer necessary. For the $s$-wave contribution, we set $J = 0$ and use $Q_0 ( 1 / v ) = ( 1 / 2 ) \ln [ ( 1 + v ) / ( 1 - v ) ]$, to find
\begin{align}
T^{( 0, 0 )}_{B; \gamma \gamma \to m^+ m^-} &= -2 i e^2 ( 1 - v^2 ) / v \nonumber \\
&\times \ln \left ( \frac{1 + v}{1 - v} \right ) \sqrt{\pi} Y_{0, 0} ( \theta, \phi ).
\label{00s}
\end{align}
The $d$-wave contribution can be obtained by setting $J = 2$ and $Q_2 ( 1 / v ) = - 3 / 2 v + ( 3 - v^2 ) / 4 v^2 \ln [ ( 1 + v ) / ( 1 - v ) ]$.

\hspace{\parindent} For the case of helicity two, since $\sin^2 \theta e^{2 i \phi} = \sqrt{32 \pi / 15} Y_{2, 2} ( \theta, \phi )$, we can write (\ref{e:bornbasis2}) as
\end{multicols}
\rule{\dimexpr(0.5\textwidth-0.5\columnsep-0.4pt)}{0.4pt}%
\begin{equation}
T^{( 2 )}_{B; \gamma \gamma \to m^+ m^-} = 2 i e^2 v \sqrt{\frac{32 \pi}{15}} \sum_{l = 0, 2, 4, \cdots} \sqrt{4 \pi} ( 2 l + 1 )^{1 / 2} Q_l ( 1 / v ) Y_{l, 0} ( \theta, \phi ) Y_{2, 2} ( \theta, \phi ).
\end{equation}
\par
\hfill
\rule[0.5\baselineskip]{\dimexpr(0.5\textwidth-0.5\columnsep-1pt)}{0.4pt}
\begin{multicols}{2}
We now make use of the expression for the product of the two spherical harmonics $Y_{l, 0} ( \theta, \phi ) Y_{2, 2} ( \theta, \phi )$ at a common angle \cite{Edmonds} to find
\end{multicols}
\rule{\dimexpr(0.5\textwidth-0.5\columnsep-0.4pt)}{0.4pt}%
\begin{align}
T^{( 2 )}_{B; \gamma \gamma \to m^+ m^-} = 2 i e^2 v &\sqrt{\frac{32 \pi}{3}} \sum_{J = 2, 4, 6, \cdots} ( 2 J + 1 )^{1 / 2} \nonumber \\
&\times \left [ \sum_{l = J - 2, J, J + 2} ( 2 l + 1 ) \left (
\begin{array}{ccc}
l & 2 & J \\
0 & 2 & -2
\end{array}
\right )
\left (
\begin{array}{ccc}
l & 2 & J \\
0 & 0 & 0
\end{array}
\right ) Q_l ( 1 / v ) \right ] Y_{J, 2} ( \theta, \phi ),
\end{align}
\par
\hfill
\rule[0.5\baselineskip]{\dimexpr(0.5\textwidth-0.5\columnsep-1pt)}{0.4pt}
\begin{multicols}{2}
where the round brackets are Wigner 3-$j$ symbols. The restriction on the sum over $l$ is due to the second Wigner 3-$j$ symbol that vanishes unless $l + 2 + J$ is even \cite{Edmonds}. Inserting their specific values given in \cite{Edmonds}, we can perform the $l$-sum in the square brackets to identify the $( J, \lambda )$ partial-wave amplitude for helicity two as
\end{multicols}
\rule{\dimexpr(0.5\textwidth-0.5\columnsep-0.4pt)}{0.4pt}%
\begin{align}
T^{( J, 2 )}_{B; \gamma \gamma \to m^+ m^-} = 4 i e^2 v &\sqrt{\pi} ( 2 J + 1 )^{1 / 2} \left [ ( J - 1 ) J ( J + 1 ) ( J + 2 ) \right ]^{1 / 2} \nonumber \\
&\times \left [ \frac{Q_{J - 2} ( 1 / v )}{( 2 J - 1 ) ( 2 J + 1 )} - 2 \frac{Q_J ( 1 / v )}{( 2 J - 1 ) ( 2 J + 3 )} + \frac{Q_{J + 2} ( 1 / v )}{( 2 J + 1 ) ( 2 J + 3 )} \right ] Y_{J, 2} ( \theta, \phi ).
\label{e:helicity2}
\end{align}

\begin{multicols}{2}
\noindent For $J = 2$, using $Q_4 ( 1 / v ) = ( -105 + 55 v^2 ) / 24 v^3$ $+ ( 35 - 30 v^2 + 3 v^4 ) / 16 v^4 \ln [ ( 1 + v ) / ( 1 - v ) ]$, we obtain
\begin{align}
&T^{( 2, 2 )}_{B; \gamma \gamma \to m^+ m^-} = 2 i e^2 \biggl [ \frac{-3 + 5 v^2}{3 v^2} + \frac{( 1 - v^2 )^2}{2 v^3} \nonumber \\
&\quad \quad \quad \quad \quad \quad \times \ln \left ( \frac{1 + v}{1 - v} \right) \biggr ] \sqrt{15 \pi / 2} Y_{2, 2} ( \theta, \phi ).
\label{e:Bornhelicity}
\end{align}

\hspace{\parindent} While the $T$-matrices will be of direct interest to us, for completeness we note that the individual helicity cross-sections can also be evaluated. The partial differential cross-section, for example, for the process $\gamma \gamma \to \pi^+ \pi^-$ reads
\begin{equation}
\frac{d \sigma^{( J, \lambda )}_B ( \gamma \gamma \to \pi^+ \pi^- )}{d \Omega} = \frac{v}{128 \pi^2 s} \left | T^{( J, \lambda )}_{B; \gamma \gamma \to \pi^+ \pi^-} \right |^2,
\end{equation}
and thus
\begin{align}
\sigma^{( J, \lambda )}_B ( \gamma \gamma \to \pi^+ \pi^- ) &= \frac{v}{128 \pi^2 s} \nonumber \\
&\quad \; \times \int_{4 \pi} d \Omega \left | T^{( J, \lambda )}_{B; \gamma \gamma \to \pi^+ \pi^-} \right |^2.
\end{align}
For the case of $( 0, 0 )$, we have
\begin{equation}
\frac{d \sigma^{( 0, 0 )}_B ( \gamma \gamma \to \pi^+ \pi^- )}{d \Omega} = \frac{v}{128 \pi^2 s} \left | T^{( 0, 0 )}_{B; \gamma \gamma \to \pi^+ \pi^-} \right |^2,
\end{equation}
so that
\begin{align}
&\sigma^{( 0, 0 )}_B ( \gamma \gamma \to \pi^+ \pi^- ) \nonumber \\
&\qquad \qquad = \frac{v}{128 \pi^2 s} \int_{4 \pi} d \Omega \left | T^{( 0, 0 )}_{B; \gamma \gamma \to \pi^+ \pi^-} \right |^2 \nonumber \\
&\qquad \qquad = \pi \frac{\alpha^2 v}{2 s} \biggl [ \frac{1 - v^2}{v} \ln \left ( \frac{1 + v}{1 - v} \right ) \biggr ]^2,
\end{align}
where $T^{( 0, 0 )}_{B; \gamma \gamma \to \pi^+ \pi^-}$ is given in (\ref{00s}). Similarly for $( 2, 2 )$, we obtain
\begin{align}
&\sigma^{( 2, 2 )}_B ( \gamma \gamma \to \pi^+ \pi^- ) \nonumber \\
&= \frac{v}{128 \pi^2 s} \int_{4 \pi} d \Omega \left | T^{( 2, 2 )}_{B; \gamma \gamma \to \pi^+ \pi^-} \right |^2 \nonumber \\
&= 15 \pi \frac{\alpha^2 v}{4 s} \biggl [ \frac{-3 + 5 v^2}{3 v^2} + \frac{( 1 - v^2 )^2}{2 v^3} \ln \left ( \frac{1 + v}{1 - v} \right ) \biggr ]^2,
\end{align}
where $T^{( 2, 2 )}_{B; \gamma \gamma \to \pi^+ \pi^-}$ is given in (\ref{e:Bornhelicity}).

\hspace{\parindent} The total cross-sections for each helicity can be computed similarly. For helicities zero and two, they read
\begin{align}
&\sigma^{( 0 )}_B ( \gamma \gamma \to \pi^+ \pi^- ) = \int_{4 \pi} d \Omega \frac{d \sigma^{( 0 )}_B}{d \Omega} \nonumber \\
&= \frac{v}{128 \pi^2 s} \int_{4 \pi} d \Omega \left | T^{( \lambda = 0 )}_{B; \gamma \gamma \to \pi^+ \pi^-} \right |^2 \nonumber \\
&= \frac{\alpha^2 v}{2 s} \int_{4 \pi} d \Omega \left ( \frac{1 - v^2}{1 - v^2 \cos^2 \theta} \right )^2 \nonumber \\
&= \pi \frac{\alpha^2 v}{s} \left [ 1 - v^2 + \frac{( 1 - v^2 )^2}{2 v} \ln \left ( \frac{1 + v}{1 - v} \right ) \right ],
\end{align}
and
\begin{align}
&\sigma^{( 2 )}_B ( \gamma \gamma \to \pi^+ \pi^- ) = \int_{4 \pi} d \Omega \frac{d \sigma^{( 2 )}_B}{d \Omega} \nonumber \\
&= \frac{\alpha^2 v}{2 s} \int_{4 \pi} d \Omega \left ( \frac{v^2 \sin^2 \theta}{1 - v^2 \cos^2 \theta} \right )^2 \nonumber \\
&= \pi \frac{\alpha^2 v}{s} \left [ 3 - v^2 - \frac{3 - 2 v^2 - v^4}{2 v} \ln \left ( \frac{1 + v}{1 - v} \right ) \right ],
\end{align}
where (\ref{e:bornbasis1}) and (\ref{e:bornbasis2}) are being used.

\hspace{\parindent} The full differential Born cross-section can then be calculated as
\begin{align}
\frac{d \sigma_B ( \gamma \gamma \to \pi^+ \pi^- )}{d \Omega} = \frac{\alpha^2 v}{2 s} &\Biggl [ \left ( \frac{1 - v^2}{1 - v^2 \cos^2 \theta} \right )^2 \nonumber \\
&+ \left ( \frac{v^2 \sin^2 \theta}{1 - v^2 \cos^2 \theta} \right )^2 \Biggr ],
\end{align}
and integrating over the full angular range yields
\begin{align}
\sigma_{B} ( \gamma \gamma \to \pi^+ \pi^- ) = 2 \pi \frac{\alpha^2 v}{s} \biggl [ 2 &- v^2 - \frac{1 - v^4}{2 v} \nonumber \\
&\times \ln \left ( \frac{1 + v}{1 - v} \right ) \biggr ],
\end{align}
which is in agreement with the result of \cite{DH93}.
\end{multicols}

\begin{multicols}{2}
\hspace{\parindent} The following remark is in order. In principle, explicit expressions for $T^{( J, 0 )}_{B; \gamma \gamma \to m^+ m^-}$ and $T^{( J, 2 )}_{B; \gamma \gamma \to m^+ m^-}$ for all allowed $J \geq \lambda$ and their corresponding partial cross-sections can be computed like those explained in this subsection. However, the leading-order partial-wave amplitudes given in (\ref{00s}) and (\ref{e:Bornhelicity}), i.e., the $( 0, 0 )$ and $( 2, 2 )$ partial waves, together account for $\sim 90 \%$ of the calculated Born cross-section for $s \lesssim 1 \; $GeV$^2$ \cite{JAOgg, Morgan} and thus build a convenient working assumption. In this regard, it is instructive to compare $\sigma_B ( \gamma \gamma \to \pi^+ \pi^- )$ with $\sigma^{( 0, 0 )}_B ( \gamma \gamma \to \pi^+ \pi^- ) + \sigma^{( 2, 2 )}_B ( \gamma \gamma \to \pi^+ \pi^- )$ as $v \to 0$. To this end, we have
\begin{align}
\sigma_B ( \gamma \gamma \to \pi^+ \pi^- ) \approx 2 \pi \frac{\alpha^2 v}{s} \bigl [ 1 &- 4 v^2 / 3 + 4 v^4 / 5 \nonumber \\
&+ \mathcal{O} ( v^6 ) \bigr ], \quad v \to 0,
\label{fb0}
\end{align}
\begin{align}
&\sigma^{( 0, 0 )}_B ( \gamma \gamma \to \pi^+ \pi^- ) + \sigma^{( 2, 2 )}_B ( \gamma \gamma \to \pi^+ \pi^- ) \nonumber \\
&\approx 2 \pi \frac{\alpha^2 v}{s} \bigl [ 1 - 4 v^2 / 3 + 32 v^4 / 45 + \mathcal{O} ( v^6 ) \bigr ], \quad v \to 0,
\label{pb0}
\end{align}
where (\ref{pb0}) slightly underestimates the full Born cross-section in (\ref{fb0}).

\subsection{Non-perturbative, strong-interaction contributions to the meson--meson reaction amplitudes}
\label{sec2b}

\hspace{\parindent} To obtain the full transition matrix shown in panel (a) of Fig.~\ref{fF1}, the effects of the strong interaction in the final state must now be taken into account. In particular, we first need to evaluate the $T$-matrices that are associated with the square diagram in Fig.~\ref{fF1} and which account for meson--meson scattering. Here, we follow the approach offered by ChPT, following \cite{JAO97}, and summarize the essential points. For recent reviews of this topic, see \cite{JAO20201, JAO20202}.

\hspace{\parindent} The meson pairs $( K \overline{K} )^I$, $( \pi \pi )^I$, and $( \pi^0 \eta )^I$ of good isospin $I$, interact in the final state via 4-point vertices given by \cite{JAO97} (note that since we define the vertex diagrams by $i V_{i j}$, the sign of this $V_{i j}$ is the negative of those listed in \cite{JAO97}),
\begin{align}
V^I_{m_1 m_2; m_3 m_4} &= \left \langle I, m_1 m_2 \left | {\cal L}_2 \right | I, m_3 m_4 \right \rangle \nonumber \\
&= V^I_{m_3 m_4; m_1 m_2},
\label{e:treelvl}
\end{align}
at tree level, after coupling both the initial and final two-meson states to good isospin $I$. The symmetry in the interaction under the interchange of labels is due to time-reversal invariance. The interaction chiral Lagrangian is assumed to be ${\cal L}_2 = 1 / 12 f^2 \text{tr} \left [ ( \partial_\mu \Phi \Phi - \Phi \partial_\mu \Phi )^2 + M \Phi^4 \right ]$, which is the leading order ChPT interaction Lagrangian density \cite{GL84, GE95}; $f$ is the (bare) pion decay constant, $\text{tr}$ denotes the trace in $SU ( 3 )$ flavor space of the matrices constructed from
\begin{equation}
\Phi = \left( \begin{smallmatrix}
\pi^0 / \sqrt 2 + \eta / \sqrt 6 & \pi^+ & K^+ \\
\pi^- & -\pi^0 / \sqrt 2 + \eta / \sqrt 6 & K^0 \\
K^- & \overline{K}^0 & -2 \eta / \sqrt 6
\end{smallmatrix} \right),
\end{equation}
and $M$ is the diagonal matrix of (bare) meson masses,
\begin{equation}
M = \left( \begin{matrix}
m^2_\pi & 0 & 0 \\
0 & m^2_\pi & 0 \\
0 & 0 & 2 m^2_K - m^2_\pi
\end{matrix} \right).
\end{equation}

\hspace{\parindent} Notationally, we abbreviate the 4-point vertices in (\ref{e:treelvl}) and the meson--meson transition amplitudes $T_{m_1 m_2 \to m_3 m_4}$ in a compact fashion as has been previously introduced in \cite{JAO97}. The indices $( i, j ) = ( 1, 2 )$ are used to identify the specific meson {\it pair} involved: 1 indicates $K \overline{K}$ in both isospin states $I = 0$ and $I = 1$, while 2 indicates $\pi \pi$ for $I = 0$ (or $I = 2$), and $\pi^0 \eta$ for $I = 1$. Note that this follows the convention of \cite{JAO97}, but is {\it opposite} to the channel-labeling convention of \cite{DaiPenn14}. The following set of basis states of good isospin $I$, $| ( M_1 M_2 )^I \rangle$, are defined in terms of the meson--meson particle-basis sets as
\begin{align}
&\left | \left ( K \overline{K} \right )^0 \right \rangle = -\frac{1}{\sqrt 2} \left ( K^+ K^- + K^0 \overline{K}^0 \right ), \nonumber \\
&\left | \left ( K \overline{K} \right )^1 \right \rangle = -\frac{1}{\sqrt 2} \left ( K^+ K^- - K^0 \overline{K}^0 \right ), \nonumber \\
&\left | \left ( \pi \pi \right )^0 \right \rangle = -\frac{1}{\sqrt 3} \left ( \pi^+ \pi^- + \pi^- \pi^+ + \pi^0 \pi^0 \right ), \nonumber \\
&\left | \left ( \pi \pi \right )^2 \right \rangle = -\frac{1}{\sqrt 6} \left ( \pi^+ \pi^- + \pi^- \pi^+ - 2 \pi^0 \pi^0 \right ), \nonumber \\
&\left | \left ( \pi^0 \eta \right )^1 \right \rangle = \pi^0 \eta.
\label{e:pi0eta}
\end{align}
\end{multicols}
\end{strip}

Using the $V^I_{i j}$, which can be obtained from (\ref{e:treelvl}) by taking (\ref{e:pi0eta}) into account, one can construct the coupled equations for the scattering amplitudes $T^I_{i j}$ of good isospin for these meson pairs. In general, these are integral equations that involve meson--meson interactions $V^I_{i j}$ in intermediate states, where at least one of the states $i$ or $j$ is off-shell. However, in the case of $s$-wave scattering, Oller and Oset have shown explicitly \cite{JAO97, JAOgg} that one can replace the $V^I_{i j}$ by their on-shell values in intermediate states too since their off-shell parts are additive and can be reabsorbed as a renormalization factor that replaces the {\it bare} coupling constant, $1 / f$, and meson masses by their physical values, $f_\pi \approx 93 \; $MeV$ $ and $( m_\pi, m_K, m_\eta ) \approx ( 140, 496, 547 ) \; $MeV$ $. The on-shell values for the $V^I_{i j}$ \footnote{Some authors, e.g., \cite{SU390, JAO97, JAOgg}, include an additional normalization of $1 / \sqrt 2$ in the definitions of $| ( \pi \pi )^0 \rangle$ and $| ( \pi \pi )^2 \rangle$. Hence the matrix elements $V^{0, 2}_{i j}$ given in Eqs.~(\ref{e28}) and (\ref{e30}) are larger by a factor $\sqrt 2$ than those listed by Oller and Oset for each pion label 2 appearing on the interaction matrix element for $I = 0$ and $I = 2$.} can then be found from the information given in \cite{JAO97, JAOgg} to be
\begin{strip}
\rule{\dimexpr(0.5\textwidth-0.5\columnsep-0.4pt)}{0.4pt}%
\begin{equation}
V^0_{1 1} = \frac{3 s}{4 f^2}, \quad V^0_{2 1} = \sqrt{\frac{3}{2}} \frac{s}{2 f^2}, \quad V^0_{2 2} = \frac{2 s - m_\pi^2}{f^2},
\label{e28}
\end{equation}
\begin{equation}
V^1_{1 1} = \frac{s}{4 f^2}, \quad V^1_{2 1} = -\sqrt{\frac{2}{3}} \frac{9 s - m_\pi^2 - 3 m_\eta^2 - 8 m_K^2}{12 f^2}, \quad V^1_{2 2} = \frac{m_\pi^2}{3 f^2},
\label{e29}
\end{equation}
\begin{equation}
V^2_{2 2} = - \frac{s - 2 m_\pi^2}{f^2},
\label{e30}
\end{equation}
\par
\hfill
\rule[0.5\baselineskip]{\dimexpr(0.5\textwidth-0.5\columnsep-1pt)}{0.4pt}
\end{strip}

\noindent where $\sqrt s$ is the total collisional energy in the center-of-mass system.

Equations (\ref{e28}) to (\ref{e30}) only depend on the external variable $s = P^2_0$, the total center-of-mass energy squared of the meson pair. This in turn means that the integration over the four-momenta of meson pairs in intermediate states can be factored \cite{JAO97}, and the coupled integral equations for the $T^I_{i j} ( s )$ become coupled algebraic equations that can be solved exactly. The results are
\begin{equation}
T^I_{1 1} ( s ) = \left [ \left ( 1 - V^I_{2 2} \Pi^I_{2 2} \right ) V^I_{1 1} + V^I_{1 2} \Pi^I_{2 2} V^I_{2 1} \right ] / D^I ( s ),
\label{e:T11}
\end{equation}
\begin{equation}
T^I_{1 2} ( s ) = V^I_{1 2} / D^I ( s ),
\label{e:T12}
\end{equation}
for $( K \overline{K} )^I \to ( K \overline{K} )^I$ in both isospin channels, $I = 0, 1$, and $( K \overline{K} )^0 \to ( \pi \pi )^0$ or $( K \overline{K} )^1 \to ( \pi^0 \eta )^1$, respectively. The common denominator of $T^I_{1 1} ( s )$ and $T^I_{1 2} ( s )$, $D^I ( s )$, reads
\begin{align}
D^I ( s ) = \bigl ( 1 - V^I_{1 1} \Pi^I_{1 1} \bigr ) \bigl ( 1 - V^I_{2 2} \Pi^I_{2 2} \bigr ) - V^I_{1 2} \Pi^I_{2 2} V^I_{2 1} \Pi^I_{1 1},
\label{e:roots}
\end{align}
for $I = 0$ and 1. In the above equations $\Pi^I_{i i}$ denotes the meson loop diagrams and is given by
\begin{equation}
i \Pi^I_{i i} ( s ) = \epsilon \int \frac{d^4 l}{( 2 \pi )^4} \frac{1}{l^2 - m^2_a} \frac{1}{( l + P_0 )^2 - m^2_b},
\label{e:Pi}
\end{equation}
for two mesons of mass $( m_a, m_b )$ as identified by the labels $I$ and $i$, and the symmetry factor $\epsilon = 1/2$ and 1 for identical and non-identical mesons propagating in the loop, respectively \cite{JLP71}. Thus $\Pi^I_{1 1} = \Pi_{K \overline{K}}$ for $K \overline{K}$ in both isospin channels, while $\Pi^I_{2 2} = \Pi_{\pi \pi}$ or $\Pi_{\pi^0 \eta}$ for $I = 0$ (or $I = 2$) and $I = 1$, respectively. Note that $T^I_{1 2} ( s ) = T^I_{2 1} ( s )$ is also time-reversal invariant.

Since pions coupled to good isospin behave like identical bosons \cite{JLP71}, $( \pi \pi )^I \to ( \pi \pi )^I$ $s$-wave scattering can only occur for $I = 0$ (or $I = 2$). The relevant $T$-matrices are
\begin{equation}
T^0_{2 2} ( s ) = \left [ T^0_{1 1} ( s ) \right ]_{1 \leftrightarrow 2},
\label{e:T22}
\end{equation}
\begin{equation}
T^2_{2 2} ( s ) = V^2_{2 2} / \left ( 1 - V^2_{2 2} \Pi^2_{2 2} \right ),
\label{e:T222}
\end{equation}
for $( \pi \pi )^0 \to ( \pi \pi )^0$ and $( \pi \pi )^2 \to ( \pi \pi )^2$, since in the latter channel only the diagonal interaction vertex $V^2_{2 2}$ is non-zero, see (\ref{e30}). The related $S$-matrix element reads $S^2_{2 2} ( s ) = 1 + i / 16 \pi \sqrt{1 - 4 m^2_\pi / s} T^2_{2 2} ( s ) = \exp ( 2 i \delta^2_2 )$, where $\delta^2_2$ is a real phase shift, $\delta^2_2 = 1 / 2 \tan^{-1} [ \text{Im} ( S^2_{2 2} ) / \text{Re} ( S^2_{2 2} ) ]$. Note that the $S$-matrix is unitary as there are no reaction channels.

The integral in (\ref{e:Pi}) diverges at large four-momenta and requires regularization. The expression for the $O ( 4 )$ regularized integral $\Pi^I_{i i} ( s )$ in (\ref{e:Pi}) depends on where $s$ lies relative to the branch cut that starts at the branch point $( m_a + m_b )^2$ \cite{SPKR11}. In the following we elaborate on the evaluation of $\Pi^I_{i i} ( s ) = \Pi ( s ) = -i \epsilon I_{a b} ( s )$.

First, we simplify $I_{a b} ( s )$ as
\begin{align}
&I_{a b} ( s ) = I_{a b} ( 0 ) - i / ( 4 \pi )^2 \nonumber \\
&\times \int_0^1 d \alpha \ln \left [ \frac{1 + ( m^2_a - m^2_b ) / m^2_b \alpha - s \alpha ( 1 - \alpha ) / m^2_b}{1 + ( m^2_a - m^2_b ) / m^2_b \alpha} \right ] \nonumber \\
&\qquad \; \; = I_{a b} ( 0 ) - \frac{i}{( 4 \pi )^2} L_{a b} ( s ).
\label{Labs1}
\end{align}
In this expression, $I_{a b} ( 0 )$ is divergent and under $O ( 4 )$ regularization, we find
\begin{align}
I_{a b} ( 0 ) = \frac{i}{( 4 \pi )^2} \biggl [ &\frac{m^2_a}{m^2_a - m^2_b} \ln \left ( 1 + \Lambda^2 / m_a^2 \right ) \nonumber \\
&- \frac{m^2_b}{m^2_a - m^2_b} \ln \left ( 1 + \Lambda^2 / m^2_b \right ) \biggr ],
\label{Iab0}
\end{align}
where $\Lambda$ is a regulatory cutoff.

To evaluate $L_{a b} ( s )$ in (\ref{Labs1}), first we consider the case of $s < 0$, for which we obtain
\begin{align}
L_{a b} ( s ) = - 1 - \frac{1}{2} \biggl ( \frac{\delta}{X} + \frac{m^2_a + m^2_b}{m^2_a - m^2_b} \biggr ) &\ln \left ( m^2_a / m^2_b \right ) \nonumber \\
&+ J_{a b} ( s ),
\end{align}
where $\delta = ( m_a - m_b ) / M$ with $M = m_a + m_b$, $X = - s / M^2$, and $J_{a b} ( s )$ is defined as
\begin{align}
J_{a b} ( s ) = \sqrt c \Biggl [ \coth^{-1} \left ( \frac{\sqrt c}{1 + \delta / X} \right ) + \coth^{-1} \left ( \frac{\sqrt c}{1 - \delta / X} \right ) \Biggr ],
\end{align}
with $c = ( 1 + \delta^2 / X ) ( 1 + 1 / X )$.

To calculate $L_{a b} ( s )$ for $s > 0$, we need to analytically continue it into the complex plane. Making the substitution $s \to z M^2$, where $z$ is a complex variable, we have
\begin{align}
J_{a b} ( z ) = \sqrt{f_{a b} ( z )} \Biggl [ &\coth^{-1} \left ( \frac{\sqrt{f_{a b} ( z )}}{1 - \delta / z} \right ) \nonumber \\
&+ \coth^{-1} \left ( \frac{\sqrt{f_{a b} ( z )}}{1 + \delta / z} \right ) \Biggr ],
\label{Jabz}
\end{align}
where $f_{a b} ( z ) = ( 1 - \delta^2 / z ) ( 1 - 1 / z )$.

\begin{figure}[!htb]

\includegraphics[scale = 1.1]{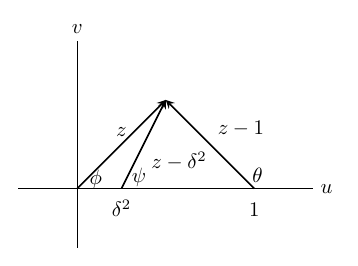}

\caption{Illustration of complex vectors defined in (\ref{comv}).}

\label{veca}
\end{figure}

To study the analytic structure of $\sqrt{f_{a b} ( z )}$, first we define
\begin{align}
z &= | z | e^{i \phi} \qquad -\pi < \phi < \pi, \nonumber \\
z - \delta^2 &= | z - \delta^2 | e^{i \psi} \qquad -\pi < \psi < \pi, \label{comv} \\
z - 1 &= | z - 1 | e^{i \theta} \qquad -\pi < \theta < \pi, \nonumber
\end{align}
where $z = u + i v$, and $| z | = \sqrt{u^2 + v^2}$, $| z - \delta^2 | = \sqrt{( u - \delta^2 )^2 + v^2}$, $| z - 1 | = \sqrt{( u - 1 )^2 + v^2}$, and the angles are defined through: $\tan \phi = v / u$, $\tan \psi = v / ( u - \delta^2 )$, and $\tan \theta = v / ( u - 1 )$; see also Fig.~\ref{veca}.

\clearpage

\begin{strip}

\begin{multicols}{2}
Thus, the individual square roots in $\sqrt{f_{a b} ( z )}$ can be written as
\begin{align}
\left ( -z \right )^{1 / 2} &= \left | z \right |^{1 / 2} e^{i \left ( \phi - \pi \right ) / 2}, \nonumber \\
\left ( \delta^2 - z \right )^{1 / 2} &= \left | \delta^2 - z \right |^{1 / 2} e^{i \left ( \psi - \pi \right ) / 2}, \label{ang} \\
\left ( 1 - z \right )^{1 / 2} &= \left | 1 - z \right |^{1 / 2} e^{i \left ( \theta - \pi \right ) / 2}, \nonumber
\end{align}
with the branch cuts taken along the real $z$-axis, $u$, from $-\infty \to 0$, $-\infty \to \delta^2$, and $-\infty \to 1$, respectively. Using (\ref{ang}), $\sqrt{f_{a b} ( z )}$ takes the form:
\begin{equation}
f^{1 / 2}_{a b} ( z ) = \left | 1 - \delta^2 / z \right |^{1 / 2} \left | 1 - 1 / z \right |^{1 / 2} e^{i \left [ \left ( \theta + \psi \right ) / 2 - \phi \right ]}.
\end{equation}

\hspace{\parindent} Just above the positive $u$-axis, we find the following forms for $f^{1 / 2}_{a b}$:
\end{multicols}
\rule{\dimexpr(0.5\textwidth-0.5\columnsep-0.4pt)}{0.4pt}%
\begin{align}
&( i ) \quad 0 < u < \delta^2: \quad \left ( \phi \to 0, \; \; \psi \to \pi, \; \; \theta \to \pi \right ) \quad f^{1 / 2}_{a b} = - \left ( \delta^2 / u - 1 \right )^{1 / 2} \left ( 1 / u - 1 \right )^{1 / 2}, \nonumber \\
&( i i ) \quad \delta^2 < u < 1: \quad \left ( \phi \to 0, \; \; \psi \to 0, \; \; \theta \to \pi \right ) \quad f^{1 / 2}_{a b} = i \left ( 1 - \delta^2 / u \right )^{1 / 2} \left ( 1 / u - 1 \right )^{1 / 2}, \label{br1} \\
&( i i i ) \quad \delta^2 < 1 < u: \quad \left ( \phi \to 0, \; \; \psi \to 0, \; \; \theta \to 0 \right ) \quad f^{1 / 2}_{a b} = \left ( 1 - \delta^2 / u \right )^{1 / 2} \left ( 1 - 1 / u \right )^{1 / 2} \nonumber,
\end{align}
\par
\hfill
\rule[0.5\baselineskip]{\dimexpr(0.5\textwidth-0.5\columnsep-1pt)}{0.4pt}
\begin{multicols}{2}
and just below the positive $u$-axis, we have
\end{multicols}
\rule{\dimexpr(0.5\textwidth-0.5\columnsep-0.4pt)}{0.4pt}%
\begin{align}
&( i ) \quad 0 < u < \delta^2: \quad \left ( \phi \to 0, \; \; \psi \to -\pi, \; \; \theta \to -\pi \right ) \quad f^{1 / 2}_{a b} = - \left ( \delta^2 / u - 1 \right )^{1 / 2} \left ( 1 / u - 1 \right )^{1 / 2}, \nonumber \\
&( i i ) \quad \delta^2 < u < 1: \quad \left ( \phi \to 0, \; \; \psi \to 0, \; \; \theta \to -\pi \right ) \quad f^{1 / 2}_{a b} = - i \left ( 1 - \delta^2 / u \right )^{1 / 2} \left ( 1 / u - 1 \right )^{1 / 2}, \label{br2} \\
&( i i i ) \quad \delta^2 < 1 < u: \quad \left ( \phi \to 0, \; \; \psi \to 0, \; \; \theta \to 0 \right ) \quad f^{1 / 2}_{a b} = \left ( 1 - \delta^2 / u \right )^{1 / 2} \left ( 1 - 1 / u \right )^{1 / 2} \nonumber.
\end{align}
\par
\hfill
\rule[0.5\baselineskip]{\dimexpr(0.5\textwidth-0.5\columnsep-1pt)}{0.4pt}
\end{strip}

\begin{figure}[!htb]

\centering

\includegraphics[scale = 0.35]{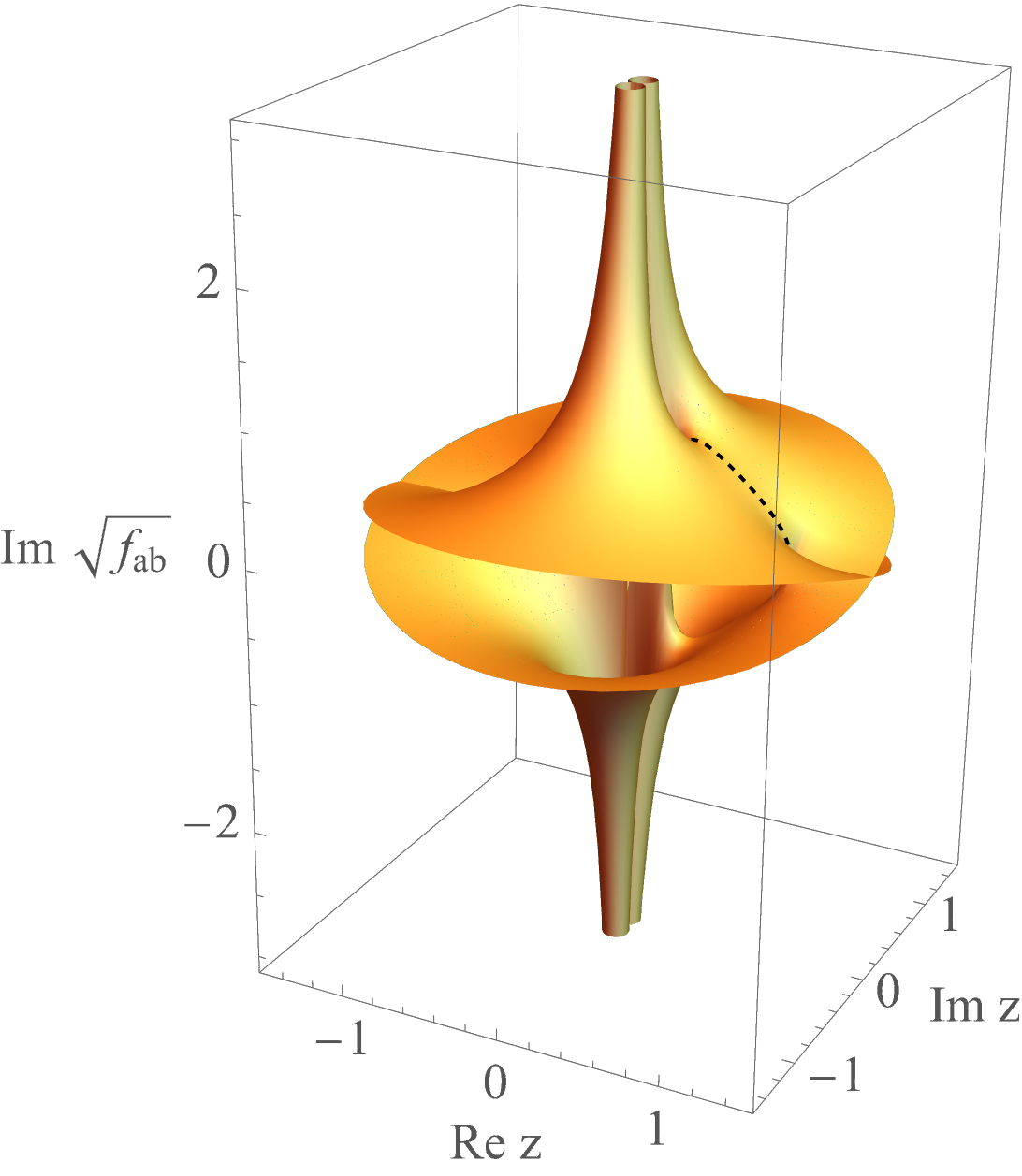}

\caption{The Riemann surface of the complex function $\sqrt{f_{a b} ( z )}$ for $\delta = 1 / 2$. The Riemann sheets are connected along the branch cut, shown as a dashed, black curve, from $1 / 4 \to 1$.}

\label{Rie}
\end{figure}

\noindent Equations~(\ref{br1}) and (\ref{br2}) demonstrate that the function $f^{1 / 2}_{a b}$ has a discontinuity along $\delta^2 \to 1$, which identifies as the branch cut of $f^{1 / 2}_{a b}$. The Riemann surface of $\sqrt{f_{a b} ( z )}$ for a specific $\delta$ is illustrated in Fig.~\ref{Rie}.

Now, we examine $J_{a b} ( z )$ in (\ref{Jabz}), in the three regions $( i )$, $( i i )$, and $( i i i )$. In both $( i )$ and $( i i )$ regions, it has the same form as
\begin{align}
J_{a b} = \sqrt{\left | f_{a b} \right |} \Bigg [ \cot^{-1} &\left ( \frac{\sqrt{\left | f_{a b} \right |}}{1 - \delta / u} \right ) \nonumber \\
&+ \cot^{-1} \left ( \frac{\sqrt{\left | f_{a b} \right |}}{1 + \delta / u} \right ) \Bigg ],
\label{reg2}
\end{align}
where $\sqrt{\left | f_{a b} \right |} = ( 1 - \delta^2 / u )^{1 / 2} ( 1 / u - 1 )^{1 / 2}$, when $u < 1$. For the region $( i i i )$, we have
\begin{align}
J_{a b} = \sqrt{\left | f_{a b} \right |} \Bigg [ &\coth^{-1} \left ( \frac{\sqrt{\left | f_{a b} \right |}}{1 - \delta / u} \right ) \nonumber \\
&+ \coth^{-1} \left ( \frac{\sqrt{\left | f_{a b} \right |}}{1 + \delta / u} \right ) \Bigg ],
\label{reg3}
\end{align}
where $\sqrt{\left | f_{a b} \right |} = ( 1 - \delta^2 / u )^{1 / 2} ( 1 - 1 / u )^{1 / 2}$, when $u > 1$. Since the arguments of both inverse hyperbolic cotangents in (\ref{reg3}) are smaller than one, therefore both lie on the upper lip---in our case, $m^2 \to m^2 - i \eta: \; z \to ( s + i \eta ) / M^2$---of the branch cut of the inverse hyperbolic cotangent function. Then (\ref{reg3}) can be written as
\begin{align}
J_{a b} = \sqrt{\left | f_{a b} \right |} \Bigg [ &\tanh^{-1} \left ( \frac{\sqrt{\left | f_{a b} \right |}}{1 - \delta / u} \right ) \nonumber \\
&+ \tanh^{-1} \left ( \frac{\sqrt{\left | f_{a b} \right |}}{1 + \delta / u} \right ) - i \pi \Bigg ].
\label{Jabz1}
\end{align}
Thus in the region $( i i i )$, where $s > M^2 = ( m_a + m_b )^2$ when the two-particle decay channel opens up, on the upper lip of the branch cut along the real axis, $J_{a b}$ develops an imaginary part.

Now with $J_{a b} ( z )$ at our disposal, the analytic continuation of $L_{a b} ( s )$ reads
\begin{align}
L_{a b} ( z ) = - 1 - \frac{1}{2} \biggl ( -\frac{\delta}{z} + \frac{m^2_a + m^2_b}{m^2_a - m^2_b} \biggr ) &\ln \left ( m^2_a / m^2_b \right ) \nonumber \\
&+ J_{a b} ( z ).
\label{Labz1}
\end{align}

Putting (\ref{Labs1}), (\ref{Iab0}), (\ref{Jabz1}), and (\ref{Labz1}) together, the regularized integral of $\Pi ( s )$ in (\ref{e:Pi}), for $s > ( m_a + m_b )^2$, becomes
\begin{strip}
\rule{\dimexpr(0.5\textwidth-0.5\columnsep-0.4pt)}{0.4pt}%
\begin{align}
\Pi ( s ) = \frac{\epsilon}{( 4 \pi )^2} \biggl [ \frac{m^2_a}{m^2_a - m^2_b} \ln \left ( 1 + \Lambda^2 / m^2_a \right ) - \frac{m^2_b}{m^2_a - m^2_b} \ln \left ( 1 + \Lambda^2 / m^2_b \right) - L_{a b} ( s ) \biggr ],
\label{e:Piclosed}
\end{align}
\par
\hfill
\rule[0.5\baselineskip]{\dimexpr(0.5\textwidth-0.5\columnsep-1pt)}{0.4pt}
\begin{multicols}{2}
with $L_{a b} ( s )$ is given by
\end{multicols}
\rule{\dimexpr(0.5\textwidth-0.5\columnsep-0.4pt)}{0.4pt}%
\begin{align}
L_{a b} ( s ) = - 1 - \frac{1}{2} \left ( \frac{m^2_a + m^2_b}{m^2_a - m^2_b} - \frac{m^2_a - m^2_b}{s} \right ) &\ln \left ( m^2_a / m^2_b \right ) + \sqrt{f_{a b}} \Biggl \{ \tanh^{-1} \Biggl [ \frac{\sqrt{f_{a b}}}{1 - \left ( m^2_a - m^2_b \right ) / s} \Biggr ] \nonumber \\
&+ \tanh^{-1} \Biggl [ \frac{\sqrt{f_{a b}}}{1 + \left ( m^2_a - m^2_b \right ) / s} \Biggr ] \Biggr \} - i \pi \sqrt{f_{a b}},
\label{labab}
\end{align}
\end{strip}
where
\begin{align}
\sqrt{f_{a b}} = &\left ( 1 - \left( m_a - m_b \right )^2 / s \right )^{1 / 2} \nonumber \\
&\times \left ( 1 - \left( m_a + m_b \right )^2 / s \right )^{1 / 2} = 2 p_{a b} / \sqrt s.
\label{e:Lab}
\end{align}
In (\ref{e:Lab}), $p_{a b}$ is the magnitude of the three-momentum of either meson in the center-of-mass system. Note that (\ref{e:Piclosed}), (\ref{labab}), and (\ref{e:Lab}) are symmetric under the interchange $m_a \to m_b$. For $m_a = m_\pi$ and $m_b = m_\eta$, (\ref{e:Piclosed}) gives the closed form of $\Pi_{\pi^0 \eta} ( s )$ for $s > ( m_\pi + m_\eta )^2$. For a different approach to calculate the meson loop diagrams, we refer to Section~3 of \cite{JAO20201}.

For equal masses, $m_a = m_b = m$, $\Pi ( s )$ reduces to
\begin{align}
\Pi ( s ) = \frac{\epsilon}{( 4 \pi )^2} \Bigl [ 1 &+ \ln \left ( 1 + \Lambda^2 / m^2 \right ) \nonumber \\
&+ m^2 / ( m^2 + \Lambda^2 ) - 2 J ( s ) \Bigr ] \label{RPi},
\end{align}
with
\begin{align}
&J ( s ) = \sqrt{- f} \cot^{-1} \sqrt{- f} \; \theta \left ( 4 m^2 - s \right ) \nonumber \\
&+ \sqrt f \left ( \tanh^{-1} \sqrt f - i \pi / 2 \right ) \theta \left ( s - 4 m^2 \right ),
\label{JS}
\end{align}
where $\sqrt f = ( 1 - 4 m^2 / s )^{1 / 2}$ and $\theta$ is the Heaviside step function. In (\ref{JS}), the analytic continuation of $J ( s )$ into the region $0 < s < 4 m^2$ along the real $s$-axis is also given. Setting $m = m_\pi$ or $m_K$ with $\epsilon = 1 / 2$ or 1, respectively, then leads to closed forms for $\Pi_{\pi \pi} ( s )$ and $\Pi_{K \overline{K}} ( s )$. Having $\Pi ( s )$ at our disposal, the $T^I_{i j} ( s )$ can all be obtained in closed form.

The complex poles in $P_0 = \sqrt s$ of the transition amplitudes $T^I_{1 1} ( P_0 )$ and $T^I_{1 2} ( P_0 )$ in (\ref{e:T11}) and (\ref{e:T12}) can be found for $I = 0$ and 1 from the roots of their common denominator $D^I ( P^I_0 ) = 0$ in (\ref{e:roots}). These roots, which determine the meson mass $M^I$ and half-width $\Gamma^I / 2$ for each isospin, lie on the appropriate second Riemann sheet in the lower half of the cut complex $P_0$-plane \cite{JAO97}, i.e., $P^I_0 = M^I - i \Gamma^I / 2$. Note that this relation assumes a non-relativistic Breit--Wigner shape for the transition amplitude in the vicinity of its peak value at $P_0 = M^I$.

One finds two roots for $I = 0$ and a single root for $I = 1$, that correspond to the two scalar-isoscalar mesons $f_0 ( 500 )$ (or $\sigma$), $f_0 ( 980 )$, and a single scalar-isovector $a_0 ( 980 )$ meson, respectively \cite{JAO97}. In the cases of $\sigma$ and $f_0$, we find particularly simple relations for their corresponding roots (in MeV) as functions of the cutoff $\Lambda$:
\begin{align}
M^0_\sigma \left ( \Lambda \right ) &\approx 436 + 67 (\Lambda / 1000) - 30 (\Lambda / 1000)^2, \nonumber \\
\Gamma^0_\sigma / 2 \left ( \Lambda \right ) &\approx 394 - 230 (\Lambda / 1000) + 50 (\Lambda / 1000)^2,
\label{s0ml}
\end{align}
and
\begin{align}
M^0_{f_0} \left ( \Lambda \right ) &\approx 996 + 37 (\Lambda / 1000) - 36 (\Lambda / 1000)^2, \nonumber \\
\Gamma^0_{f_0} / 2 \left ( \Lambda \right ) &\approx -64 + 93 (\Lambda / 1000) - 21 (\Lambda / 1000)^2,
\label{f0ml}
\end{align}
where $\Lambda$ is measured in MeV and the coefficients of $\Lambda^0$, $\Lambda$, and $\Lambda^2$ have dimensions of MeV, 1, and MeV$^{-1}$, respectively. We use the $O ( 4 )$ cutoff of $\Lambda = 1351^{+160}_{-185} \; $MeV$ $ that fixes the real part of one $I = 0$ root, with error bars, at $M^0_{f_0} = ( 980 \pm 10 ) \; $MeV$ $ \cite{SPKR11}. This replicates the $f_0 ( 980 )$ mass values quoted in the PDG data table \cite{PDG10}. The predictions for the masses and half-widths for all three scalar mesons are listed in Table~\ref{t:table1}.
\begin{table*}[!htb]

\centering

\caption[]{Summary of calculated masses and half-widths, $M^I - i \Gamma^I / 2$ (in MeV, rounded to the nearest integer), for $f_0 ( 500 )$ or $\sigma$, $f_0 ( 980 )$, and $a_0 ( 980 )$ as a function of the range of $O ( 4 )$ cutoffs: $\Lambda = 1351^{+160}_{-185} \; $MeV$ $. The other input parameters are taken from experiment: $f_\pi = 93 \; $MeV$ $ and $( m_\pi, m_K, m_\eta ) = ( 140, 496, 547 ) \; $MeV$ $.}
\label{t:table1}
\renewcommand{\arraystretch}{2.0}
\begin{tabular}{@{}llll@{}}
\toprule
$\Lambda \; $MeV$ $ & $M^0_\sigma - \frac{i}{2} \Gamma^0_\sigma$ & $M^0_{f_0} - \frac{i}{2} \Gamma^0_{f_0}$ & $M^1_{a_0} - \frac{i}{2} \Gamma^1_{a_0}$ \\
\midrule
$1166$ & $473 - 194 i$ & $990 - 16 i$ & $987 - 52 i$ \\
$1351$ & $472 - 175 i$ & $980 - 23 i$ & $964 - 53 i$ \\
$1511$ & $469 - 161 i$ & $970 - 29 i$ & $946 - 52 i$ \\
\midrule
PDG (2010) \cite{PDG10} & $( 400 - 1200 ) - ( 300 - 500 ) i$ & $( 980 \pm 10 ) - ( 20 - 50 ) i$ & $( 980 \pm 20 ) - ( 25 - 50 ) i$ \\
PDG (2012) \cite{PDG10} & $( 400 - 550 ) - ( 200 - 350 ) i$ & $( 990 \pm 20 ) - ( 20 - 50 ) i$ & $( 980 \pm 20 ) - ( 25 - 50 ) i$ \\
PDG (2022) \cite{PDG10} & $( 400 - 550 ) - ( 200 - 350 ) i$ & $( 980 - 1010 ) - ( 20 - 35 ) i$ & $( 960 - 1030 ) - ( 20 - 70 ) i$ \\
\botrule
\end{tabular}
\end{table*}

For other approaches to determine the pole positions of $f_0(980)$ and $a_0(980)$ resonances, such as using the Roy-like GKPY equations and Flatt\'e parametrization, Madrid-Krakow dispersive parametrization, and unitarization techniques based on $N/D$ method, we refer to \cite{SMR1971, RGM2011, ZQW2022, VB2023, JAO1999, ZHG2011, ZHG20121, ZHG20122}.

\newpage

We are now able to construct the contribution in which the strong interaction influences the final state. According to the third diagram in panel (a) of Fig.~\ref{fF1}, we identify $T_{S; \gamma \gamma \to m_1 m_2}^{\mu \nu}$ for a propagating charged meson pair $m_+ m_-$ in the loop as
\begin{strip}
\rule{\dimexpr(0.5\textwidth-0.5\columnsep-0.4pt)}{0.4pt}%
\begin{align}
T_{S; \gamma \gamma \to m_1 m_2}^{\mu \nu} &= \int \frac{d^4 l}{\left ( 2 \pi \right )^4} \left \langle q_1, q_2 \left | T_B^{\mu \nu} \right | q_1 + l, q_2 - l \right \rangle \left \langle q_1 + l, q_2 - l \left | T_{m_1 m_2; m_+ m_-} \right | p_1, p_2 \right \rangle \nonumber \\
&\approx T_{m_1 m_2; m_+ m_-} ( s ) \int \frac{d^4 l}{\left ( 2 \pi \right )^4} \left \langle q_1, q_2 \left | T_B^{\mu \nu} \right | q_1 + l, q_2 - l \right \rangle,
\label{ts}
\end{align}
\par
\hfill
\rule[0.5\baselineskip]{\dimexpr(0.5\textwidth-0.5\columnsep-1pt)}{0.4pt}
\begin{multicols}{2}
with $m_{\pm}$ being the charged meson mass $m_\pi$ or $m_K$, as appropriate. The second line in (\ref{ts}) follows after factoring out the meson--meson scattering box diagram from the integral. This approximation, like that for $T^I_{i j} ( s )$, places the intermediate incoming charged meson pair $m_+ m_-$ of $T_{m_1 m_2; m_+ m_-}$ on-shell to render this amplitude a function of $s$ only \cite{SPKR11, JAO97, JAOgg}. Note, however, that the charged mesons in the final state of $T^{\mu \nu}_B$ under the integral sign are, by contrast, both still off-shell.

\hspace{\parindent} Equation~(\ref{ts}) can be further calculated as
\begin{align}
&T_{S; \gamma \gamma \to m_1 m_2}^{\mu \nu} = - \frac{e^2}{2 \pi^2} \left [ g^{\mu \nu} \left ( q_1 \cdot q_2 \right ) - q^\mu_2 q^\nu_1 \right ] \nonumber \\
&\times \frac{J_{m_\pm} ( s )}{2 s} T_{m_1 m_2; m_+ m_-} ( s ),
\label{gau}
\end{align}
where the function $J_{m_\pm} ( s )$ is given by
\begin{align}
J_{m_\pm} ( s ) = &1 + 2 m^2_{\pm} / s \nonumber \\
&\times \int_0^1 \frac{d \alpha}{\alpha} \ln \left [ 1 - \frac{s}{m^2_{\pm}} \alpha \left ( 1 - \alpha \right ) \right ].
\label{jif}
\end{align}
The meson--meson scattering contribution, i.e., $T_{m_1 m_2; m_+ m_-} ( s )$, can be constructed from (\ref{e:T11}) to (\ref{e:T222}) for the appropriate meson pairs. The expression in (\ref{gau}) is fully gauge invariant as the factor $g^{\mu \nu} ( q_1 \cdot q_2 ) - q^\mu_2 q^\nu_1$ guarantees this: $q_{1 \mu} [ g^{\mu \nu} ( q_1 \cdot q_2 ) - q^\mu_2 q^\nu_1 ] = 0$ and $q_{2 \nu} [ g^{\mu \nu} ( q_1 \cdot q_2 ) - q^\mu_2 q^\nu_1 ] = 0$.

\hspace{\parindent} The function $J_{m_\pm} ( s )$ can be further evaluated to
\end{multicols}
\begin{align}
J_{m_\pm} ( s ) = \left [ 1 - \frac{4 m^2_\pm}{s} \left ( \sin^{-1} \sqrt{\frac{s}{4 m^2_\pm}} \right )^2 \right ] \theta \left ( 4 m^2_\pm - s \right ) + \Biggl [ 1 + \frac{4 m^2_\pm}{s} \biggl ( \cosh^{-1} \sqrt{\frac{s}{4 m^2_\pm}} -i \pi / 2 \biggr )^2 \Biggr ] \theta \left ( s - 4 m^2_\pm \right ).
\label{e:JpiK}
\end{align}
\par
\hfill
\rule[0.5\baselineskip]{\dimexpr(0.5\textwidth-0.5\columnsep-1pt)}{0.4pt}
\end{strip}

\noindent In this expression, the second term for $s \geq 4 m^2_\pm$ arises by analytically continuing the first as a function of $s$ onto the upper lip of the branch cut $4 m^2_\pm < s < \infty$ along the real axis of the complex $s$-plane.

Allowing for different combinations of intermediate meson pairs to be formed, the contribution from the final-state strong interactions to the full contracted $T$-matrix thus reads
\begin{align}
&T_{S; \gamma \gamma \to m_1 m_2} = \epsilon_{2 \mu} T_{S; \gamma \gamma \to m_1 m_2}^{\mu \nu} \epsilon_{1 \nu} \nonumber \\
&= - \frac{e^2}{8 \pi^2} \epsilon_2 \cdot \epsilon_1 \sum_{m_\pm} J_{m_\pm} ( s ) T_{m_1 m_2; m_+ m_-} ( s ),
\label{e:FSI}
\end{align}
and when contracted with respect to polarization vectors of total helicity $\lambda$, we have
\begin{align}
&\epsilon_{i \lambda'} ( 2 ) T_{S; \gamma \gamma \to m_1 m_2}^{i j} \epsilon_{j \lambda} ( 1 ) = \frac{e^2}{8 \pi^2} {\bf e}_{\lambda^{'}} ( 2 ) \cdot {\bf e}_\lambda ( 1 ) \nonumber \\
&\qquad \qquad \quad \times \sum_{m_\pm} J_{m_\pm} ( s ) T_{m_1 m_2; m_+ m_-} ( s ).
\label{e:F00}
\end{align}
This scattering amplitude is independent of the scattering angle and is thus pure $s$-wave. Hence
\begin{align}
T^{( 0, 0 )}_{S; \gamma \gamma \to m_1 m_2} = \frac{e^2}{8 \pi^2} \sum_{m_\pm} J_{m_\pm} ( s ) &T_{m_1 m_2; m_+ m_-} ( s ) \nonumber \\
&\times 2 i \sqrt \pi Y_{0, 0}.
\label{e:FSIhel}
\end{align}
Note that there is no $\lambda = 2$ contribution in this case.

\subsection{Non-chiral perturbation theory contributions}
\label{sec2c}

As already commented upon in the Introduction, additional $s$ channel background contributions from the $f_2 ( 1270 )$ and $a_2 ( 1320 )$ resonances, with quantum numbers $I^G ( J^{P C} ) = 0^+ ( 2^{+ +} )$ and $1^- ( 2^{+ +} )$, to the scattering amplitude can be expected to be important at center-of-mass energies below 1 GeV. We denote these as $T_{R; \gamma \gamma \to m_1 m_2}^{( 2, 2 )}$. Such contributions lie beyond the scope of the ChPT calculations outlined above and are thus parametrized. In addition, the $t$-channel axial exchange amplitude arising from the $1^- ( 1^{+ +} )$ $a_1 ( 1260 )$ resonance, denoted as $T_{A; \gamma \gamma \to \pi^0 \eta}$, also not only plays a role above $\sim {\cal O} ( 1 \; $GeV$ )$ \cite{JAOgg, DH93}, but also influences the amplitudes and cross-sections below 1 GeV.

The $f_2$ and $a_2$ resonances have been interpreted as pure $d$-wave, helicity 2 states $( J, \lambda ) = ( 2, 2 )$ \cite{Albrecht89}. We parametrize these by a relativistic Breit--Wigner resonance amplitude \cite{BWagner87} for $\gamma \gamma \to M_R \to m_1 m_2$ as
\begin{equation}
T_{R; \gamma \gamma \to m_1 m_2}^{( 2, 2 )} = -16 \pi i \sqrt{20 \pi / v} A_{2, 2} ( s ) Y_{2, 2} ( \theta, \phi ),
\label{e:resonanceamp}
\end{equation}
where
\begin{align}
A_{2, 2} ( s ) &= \left[ \Gamma^{( 2 )}_{\gamma \gamma} \right]^{1 / 2} \text{BW} ( s ) \nonumber \\
&= \left[ \Gamma^{( 2 )}_{\gamma \gamma} \right]^{1 / 2} \frac{\sqrt s}{s - M^2_R + i M_R \Gamma ( \sqrt s )} \nonumber \\
&\qquad \qquad \quad \; \; \; \; \; \; \; \times \left[ \text{Br} ( m_1 m_2 ) \Gamma ( \sqrt s ) \right]^{1 / 2},
\end{align}
with $R = f_2, a_2$, for a resonance of mass and total width $M_R$ and $\Gamma ( \sqrt s )$, respectively, and with partial widths $\Gamma^{( 2 )}_{\gamma \gamma}$ and $\text{Br} ( m_1 m_2 ) \Gamma ( \sqrt s )$ for the decay into two photons of opposite helicity, or two mesons, respectively; $\text{Br} ( m_1 m_2 )$ is the branching ratio for the latter decay. For simplicity, we consider the widths appearing in the relativistic Breit--Wigner formula to be independent of $s$, i.e., $\Gamma ( \sqrt s ) = \Gamma$.

To obtain the corresponding cross-section, first note that
\begin{equation}
\frac{d \sigma^{( 2, 2 )}_R ( \gamma \gamma \to m_1 m_2 )}{d \Omega} = \frac{v}{128 \pi^2 s} \left | T^{( 2, 2 )}_{R; \gamma \gamma \to m_1 m_2} \right |^2,
\end{equation}
thus, we have
\begin{align}
&\frac{d \sigma^{( 2, 2 )}_R ( \gamma \gamma \to m_1 m_2 )}{d \Omega} = \frac{40 \pi}{s} \left | A_{2, 2} ( s ) \right |^2 \left | Y_{2, 2} ( \theta, \phi ) \right |^2 \nonumber \\
& = 40 \pi \Gamma^{( 2 )}_{\gamma \gamma} \frac{1}{( s - M^2_R )^2 + ( M_R \Gamma )^2} \text{Br} ( m_1 m_2 ) \Gamma \nonumber \\
&\qquad \qquad \qquad \qquad \qquad \qquad \qquad \times \left | Y_{2, 2} ( \theta, \phi ) \right |^2.
\end{align}
By integrating over the full angular range, we obtain
\begin{align}
\sigma^{( 2, 2 )}_R ( \gamma \gamma \to m_1 m_2 ) &= 40 \pi \Gamma^{( 2 )}_{\gamma \gamma} \text{Br} ( m_1 m_2 ) \Gamma \nonumber \\
&\quad \times \frac{1}{( s - M^2_R )^2 + ( M_R \Gamma )^2},
\end{align}
which peaks at $s = M_R^2$ with the maximum value of $40 \pi \Gamma^{( 2 )}_{\gamma \gamma} \text{Br} ( m_1 m_2 ) / M_R^2 \Gamma$.

Due to their large total widths ($\sim 100$ to $200$ MeV), the $f_2$ and $a_2$ resonances can contribute to production cross-sections already at energies $\sim 1 \; $GeV$ $, well below their peak positions. We illustrate this in the next section for the $\gamma \gamma \to \pi^0 \pi^0$ channel where the total cross-section, (\ref{e:fullcross}), is just the sum of the partial cross-sections determined by ChPT and the $f_2$ resonance amplitudes separately, without any interference term.

For the axial vector resonance exchange contribution, we follow the approach of \cite{JAOgg}, where for $\gamma(q_1) \gamma(q_2) \to p_1 p_2$, the corresponding transition amplitude reads
\begin{align}
&T_A^{\mu \nu} = 4 \pi \alpha \frac{f_A^2}{f_{\pi}^2} \left[ g^{\mu \nu} \left( q_1 \cdot q_2 \right) - q_2^\mu q_1^\nu \right] \nonumber \\
&\times \left( \frac{1 + p_1 (q_1 - p_1)/m_A^2}{(q_1 - p_1)^2 - m_A^2} + \frac{1 + p_1 (q_2 - p_1)/m_A^2}{(q_2 - p_1)^2 - m_A^2} \right),
\end{align}
where $f_A$ depends on the combination $L_9 + L_{10}$.

\subsection{Total $\gamma \gamma \to$ meson--meson transition amplitudes and cross-sections}
\label{sec2d}

The total contracted transition amplitudes $T_{\gamma \gamma \to m_1 m_2}$ can now be calculated from (\ref{e:tmunu}). For each specific exit channel, this can be written to leading order as a sum of $s$- and $d$-wave components:
\begin{equation}
T_{\gamma \gamma \to m_1 m_2} = T^{( 0, 0 )}_{\gamma \gamma \to m_1 m_2} + T^{( 2, 2 )}_{\gamma \gamma \to m_1 m_2},
\label{e:tdecompose}
\end{equation}
where
\begin{align}
T^{( J, \lambda )}_{\gamma \gamma \to m_1 m_2} = T^{( J, \lambda )}_{B; \gamma \gamma \to m_1 m_2} &+ T^{( J, \lambda )}_{S; \gamma \gamma \to m_1 m_2} \nonumber \\
&+ T^{( J, \lambda )}_{R ( A ); \gamma \gamma \to m_1 m_2},
\label{e:dec2}
\end{align}
and the relevant components are selected for the Born and strong interaction terms from ChPT, while resonant, non-ChPT terms are parametrized.

The total $\gamma \gamma \to$ meson--meson cross-section can then be expressed as the sum of the moduli squared of the $T^{( J, \lambda )}$, due to the orthogonality of the spherical harmonics they contain. Thus, we have

\begin{strip}
\rule{\dimexpr(0.5\textwidth-0.5\columnsep-0.4pt)}{0.4pt}%
\begin{align}
\sigma ( \gamma \gamma \to m_1 m_2 ) &= \frac{v}{128 \pi^2 s} \epsilon \int d \Omega \left| T_{\gamma \gamma \to m_1 m_2} \right|^2 = \frac{v}{128 \pi^2 s} \epsilon \int d \Omega \left( \left| T^{( 0, 0 )}_{\gamma \gamma \to m_1 m_2} \right|^2 + \left| T^{( 2, 2 )}_{\gamma \gamma \to m_1 m_2} \right|^2 \right) \nonumber \\
&= \sigma^{( 0, 0 )} ( \gamma \gamma \to m_1 m_2 ) + \sigma^{( 2, 2 )} ( \gamma \gamma \to m_1 m_2 ),
\label{e:fullcross}
\end{align}
\par
\hfill
\rule[0.5\baselineskip]{\dimexpr(0.5\textwidth-0.5\columnsep-1pt)}{0.4pt}
\end{strip}

\noindent after integrating over the full solid angle and averaging over the two helicities of the incoming photon pair; $\epsilon$ takes $1 / 2$ or 1, depending on whether the final state has identical particles or not, respectively. For example, for the process $\gamma \gamma \to \pi^+ \pi^-$, we obtain the total cross-section as
\begin{align}
\sigma ( \gamma \gamma \to \pi^+ \pi^- ) = &\sigma^{( 0, 0 )}_{B + S} ( \gamma \gamma \to \pi^+ \pi^- ) \nonumber \\
&\qquad + \sigma^{( 2, 2 )}_{B + f_2} ( \gamma \gamma \to \pi^+ \pi^- ),
\end{align}
where the subscripts $B + S$ and $B + f_2$ refer to the Born plus strong-interaction and Born plus resonant $f_2 ( 1270 )$ contributions, respectively.

The individual $T_{\gamma \gamma \to m_1 m_2}$ matrices, $T^{( 0, 0 )}_{\gamma \gamma \to m_1 m_2}$ and $T^{( 2, 2 )}_{\gamma \gamma \to m_1 m_2}$, that determine these partial cross-sections are given explicitly in Table~\ref{t:table2} for the three exit channels: $m_1 m_2 = \pi^0 \pi^0, \; \pi^+ \pi^-$, and $\pi^0 \eta$.
\begin{sidewaystable}

\centering

\caption[]{Combinations of partial $T$-matrices of good angular momentum and helicity, $( J, \lambda )$, that determine $| T_{\gamma \gamma \to m_1 m_2} |^2$ in Eq.~(\ref{e:fullcross}) for the production cross-section of a meson pair, $\gamma \gamma \to m_1 m_2$. The contributions to the $T$-matrices incorporating the additional resonant and $t$-channel axial exchange contributions, not generated by the ChPT Lagrangian, are labeled explicitly by the resonance or axial term that characterizes them.}

\label{t:table2}

\renewcommand{\arraystretch}{2.0}

\begin{tabular}{@{}llll@{}}
\toprule
$ m_1 m_2$ & $\epsilon$ & $v = p / q$ & $\left| T_{\gamma \gamma \to m_1 m_2} \right|^2 = \left| T^{( 0, 0 )}_{\gamma \gamma \to m_1 m_2} \right|^2 + \left| T^{( 2, 2 )}_{\gamma \gamma \to m_1 m_2} \right|^2$ \\
\midrule
$\pi^0 \pi^0$ & $1 / 2$ & $\left( 1 - \frac{4 m_\pi^2}{s} \right)^{1 / 2}$ & $\left| T^{( 0, 0 )}_{S; \gamma \gamma \to \pi^0 \pi^0} \right|^2 + \left| T^{( 2, 2 )}_{f_2; \gamma \gamma \to \pi^0 \pi^0} \right|^2$ \\
$\pi^+ \pi^-$ & 1 & $\left( 1 - \frac{4 m_\pi^2}{s} \right)^{1 / 2}$ & $\left| T^{( 0, 0 )}_{B; \gamma \gamma \to \pi^+ \pi^-} + T^{( 0, 0 )}_{S; \gamma \gamma \to \pi^+ \pi^-} \right|^2 + \left| T^{( 2, 2 )}_{B; \gamma \gamma \to \pi^+ \pi^-} + T^{( 2, 2 )}_{f_2; \gamma \gamma \to \pi^+ \pi^-} \right|^2$ \\
$\pi^0 \eta$ & 1 & $\left[ 1 - \frac{( m_\pi - m_\eta )^2}{s} \right]^{1 / 2} \left[ 1 - \frac{( m_\pi + m_\eta )^2}{s} \right]^{1 / 2}$ & $\left| T^{( 0, 0 )}_{S; \gamma \gamma \to \pi^0 \eta} + T^{( 0, 0 )}_{A; \gamma \gamma \to \pi^0 \eta} \right|^2 + \left| T^{( 2, 2 )}_{a_2; \gamma \gamma \to \pi^0 \eta} \right|^2$ \\
\botrule
\end{tabular}

\end{sidewaystable}

\section{Numerical results for scattering amplitudes and cross-sections}
\label{sec3}

\subsection{$m_1 m_2 \to m_3 m_4$ scattering amplitudes}

\begin{figure}

\begin{center}
\includegraphics[scale = 0.50]{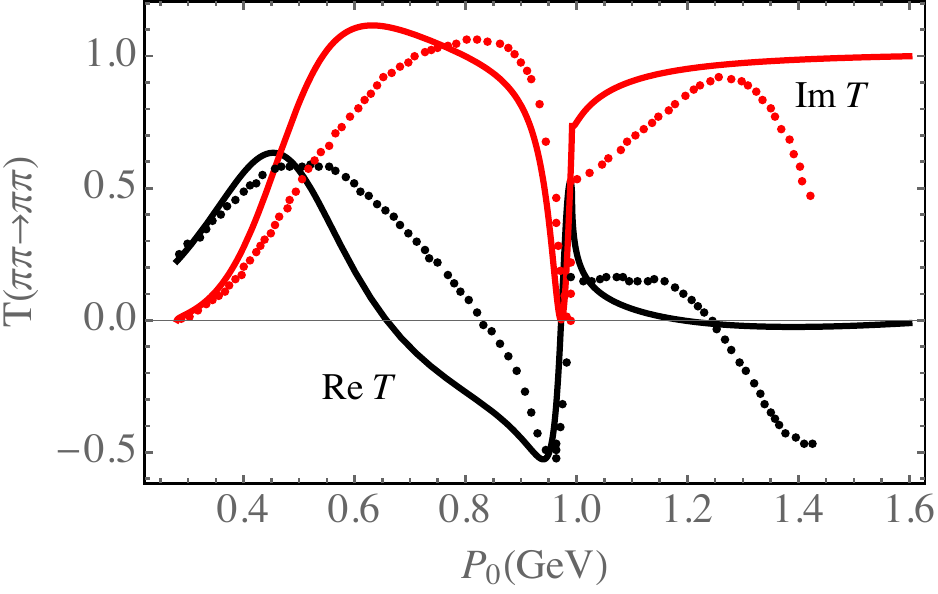}
\hspace{0.3cm}
\includegraphics[scale = 0.50]{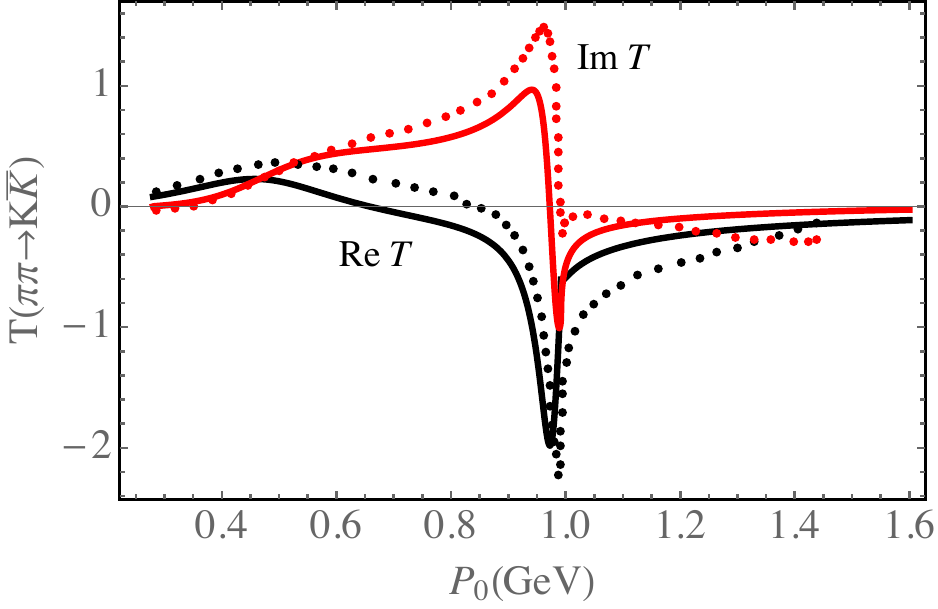}
\hspace{0.3cm}
\includegraphics[scale = 0.50]{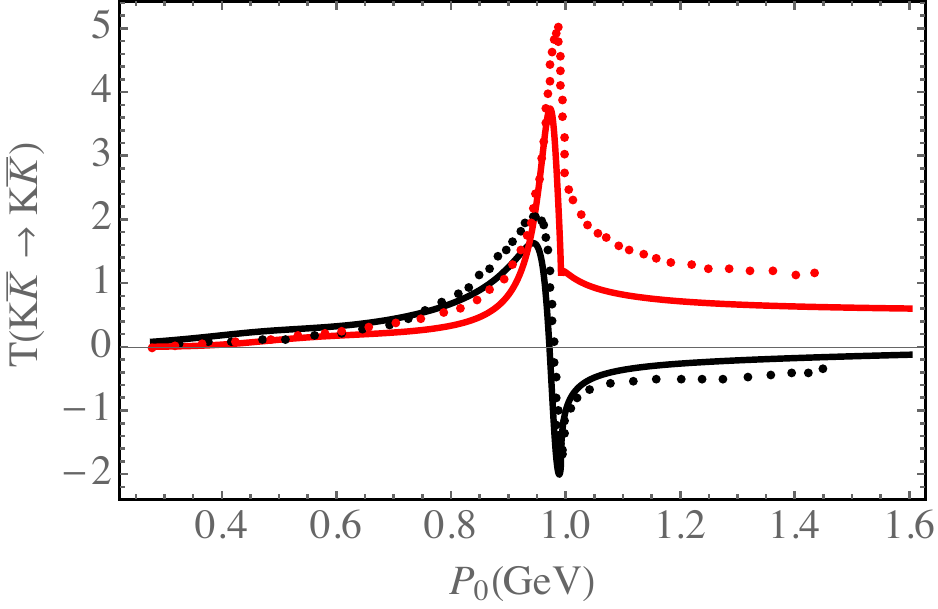}
\end{center}

\caption{The $T$-matrix elements for the meson--meson scattering amplitudes of $\pi \pi \to \pi \pi$, $\pi \pi \to K \overline{K}$, and $K \overline{K} \to K \overline{K}$. The solid curves are our calculations and the dotted curves are the values extracted from the data by Dai and Pennington \cite{DaiPenn14}. The real parts are given by the black curves and the imaginary values by the red ones.}

\label{F2}
\end{figure}

One of the main contributions to the total scattering amplitude in (\ref{e:tmunu}) arises from final-state strong interactions through meson--meson scattering. The latter amplitudes themselves are an important input to the photon--photon cross-sections and are calculated separately in Sec.~\ref{sec2b} within the framework of ChPT. The authors of \cite{DaiPenn14} have extracted these amplitudes from the data in a model-independent fashion. Here, we compare our calculated results for the real and imaginary parts of the transition matrices with the results of their fits, see Fig.~\ref{F2}, for the processes $\pi \pi \to \pi \pi$, $\pi \pi \to K \overline{K}$, and $K \overline{K} \to K \overline{K}$.
As can be seen in this figure, there is an overall good qualitative agreement between the transition matrices calculated from ChPT and those extracted from experiment. However, quantitatively there are differences, notably for the transition $\pi \pi \to \pi \pi$, in which the real part underestimates the extracted values at energies below 1 GeV, while the imaginary part overestimates the extracted values in the lower energy range, peaking at a lower value of $P_0$.

To conclude this subsection, we note that the amplitudes calculated here can also give us the effective coupling constants responsible for the decay modes of $f_0$. To see this, first note that the $f_0$ propagator, $-i D(s)$, required to construct the $T$-matrices for $I = 0$ scattering of $K \overline{K}$ via $s$ channel $f_0$ exchange in the vicinity of the $f_0$ resonance, is dressed by proper polarization loops, $-i \Pi(s)$, in the ladder sum as
\begin{align}
-i D(s) &= \frac{i}{s - M^2} + \frac{i}{s - M^2} \frac{1}{i} \Pi(s) \frac{i}{s - M^2} + \cdots \nonumber \\
&= \frac{i}{s - M^2 - \Pi(s)},
\end{align}
where $M$ is the \textit{bare} mass of $f_0$. Now, using the Lagrangian fragment, $\delta \mathcal L$, we can define the effective coupling constants via
\begin{equation}
\delta \mathcal L = g_{f_0 K \overline{K}} f_0(x) K \overline{K} + g_{f_0 \pi \pi} f_0(x) \pi \pi.
\label{fl}
\end{equation}
This Lagrangian leads to 3-point vertices given by $i V_{f_0 K^+ K^-} = i V_{f_0 K^0 \overline{K}^0} = i g_{f_0 K \overline{K}}$. Since the kaon--antikaon state of isospin $I = 0$ is $|(K \overline{K})^0 \rangle = -1 / \sqrt 2 (K^+ K^- + K^0 \overline{K}^0)$, thus the $K \overline{K}$ coupling vertex with the isoscalar $f_0$ is $i V^0_{f_0 K \overline{K}} = -\sqrt 2 i g_{f_0 K \overline{K}}$. Similarly, we can obtain $i V_{f_0 \pi^+ \pi^-} = i V_{f_0 \pi^- \pi^+} = i V_{f_0 \pi^0 \pi^0} = 2 i g_{f_0 \pi \pi}$ and the associated coupling vertex of the two-pion state of $I = 0$ reads $i V^0_{f_0 \pi \pi} = 2 \sqrt 3 i g_{f_0 \pi \pi}$.

\begin{figure}[!htb]

\includegraphics[scale = 0.55]{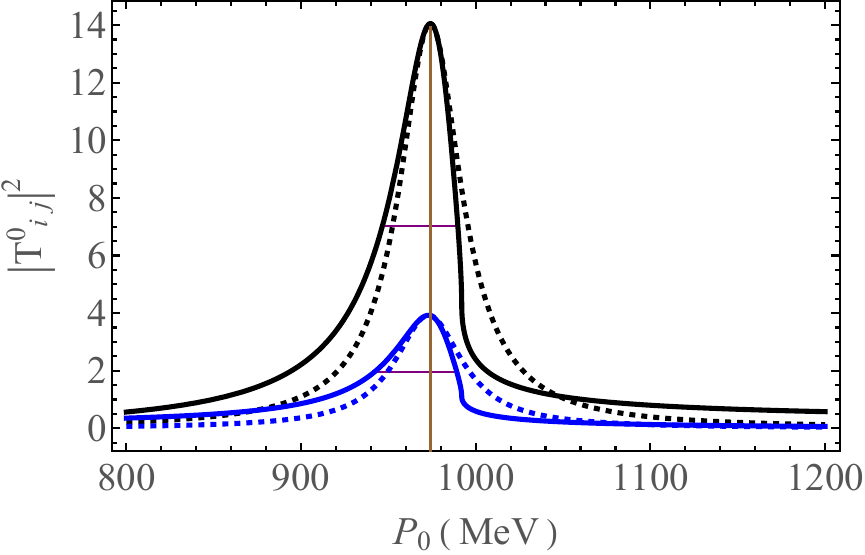}

\caption{The solid curves depict the ChPT-generated $|T^0_{1 1}(s)|^2$ (black) and $|T^0_{2 1}(s)|^2$ (blue). Their Breit--Wigner fits are represented as dashed curves, accordingly. The horizontal bars (purple) indicate the full widths at half-maximum for the $|T^0_{i j}(s)|^2$; the vertical line (brown) indicates the (almost) common value of their resonance energy at $\approx 974 \; $MeV$ $.}

\label{TijP0}
\end{figure}

\begin{figure}[!htb]

\includegraphics[scale = 0.55]{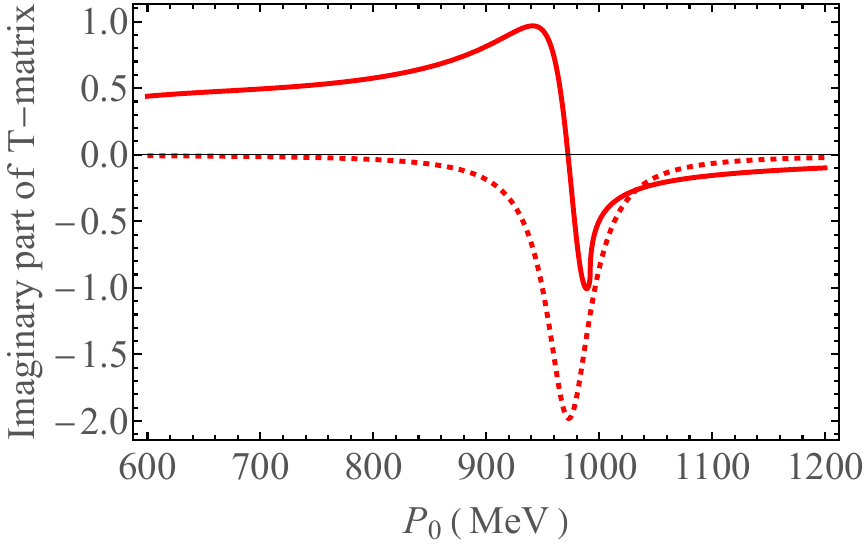}

\caption{Comparison of the exact (ChPT) $\text{Im} \; T^0_{2 1} (s)$ (solid curve) with its Breit--Wigner approximation (dashed curve). The figure illustrates that the coupling constant product $g_{f_0 \pi \pi} g_{f_0 K \overline{K}}$ has to be negative.}

\label{ImT}
\end{figure}

If we construct $-i \Pi(s)$, using the second term in (\ref{fl}), together with $i V^0_{f_0 \pi \pi}$ for the coupling constant in the $I = 0$ channel, then we can write $\Pi(s) = \text{Re} \; \Pi(s) + i \text{Im} \; \Pi(s)$, with
\begin{align}
-\text{Im} \; \Pi(s) &= \frac{3 g_{f_0 \pi \pi}^2}{8 \pi} \left( 1 - \frac{4 m_{\pi}^2}{s} \right)^{1 / 2} \theta \left( s - 4 m_{\pi}^2 \right) \nonumber \\
&= \sqrt s \Gamma(s), \label{ImPi}
\end{align}
where we have used (\ref{e:Pi}), (\ref{RPi}), and (\ref{JS}). Equation (\ref{ImPi}) can also be obtained using Cutkosky rules for cutting a loop integral, see \cite{RHL09, LDL1982}. Note that the imaginary part of $\Pi(s)$ is independent of the mode regularization. We now absorb the real part of $\Pi(s)$ in the bare mass to give the physical mass of $f_0$, i.e., $M^2 + \text{Re} \; \Pi \to M_0^2$, and evaluate $\text{Im} \; \Pi(s)$ at this physical mass, $s = M_0^2$, to define the decay width $\Gamma_0$ into two pions via $-\text{Im} \; \Pi(M_0^2) = M_0 \Gamma_0$ as
\begin{equation}
M_0 \Gamma_0 = \frac{3 g_{f_0 \pi \pi}^2}{8 \pi} \left( 1 - \frac{4 m_{\pi}^2}{M_0^2} \right)^{1 / 2} \theta \left( M_0^2 - 4 m_{\pi}^2 \right).
\end{equation}
Now using the 3-point vertices expressed in terms of the coupling constants, the $T$-matrices of $K \overline{K} \to K \overline{K}$ and $K \overline{K} \to \pi \pi$ can be written as
\begin{align}
T_{K \overline{K} \to K \overline{K}} &= (\sqrt 2 i g_{f_0 K \overline{K}})^2 \frac{1}{s - M_0^2 + i M_0 \Gamma_0} \nonumber \\
&\approx \frac{1}{2 M_0} (\sqrt 2 i g_{f_0 K \overline{K}})^2 \frac{1}{P_0 - M_0 + i \Gamma_0 / 2},
\end{align}
and
\begin{align}
&T_{K \overline{K} \to \pi \pi} = (\sqrt 2 i g_{f_0 K \overline{K}}) (2 \sqrt 3 i g_{f_0 \pi \pi}) \nonumber \\
&\qquad \qquad \quad \times \frac{1}{s - M_0^2 + i M_0 \Gamma_0} \nonumber \\
&\approx \frac{1}{2 M_0} (\sqrt 2 i g_{f_0 K \overline{K}}) (2 \sqrt 3 i g_{f_0 \pi \pi}) \frac{1}{P_0 - M_0 + i \Gamma_0 / 2}.
\end{align}
Since the $f_0$ mass corresponds to the peak position of the $|T^0_{1 1} (s)|^2$, see also Sec.~\ref{sec2b}, we can obtain the coupling constants by fitting a Breit--Wigner to $T^0_{1 1} (s) \to T_{K \overline{K} \to K \overline{K}}$ and to $T^0_{2 1} (s) \to T_{K \overline{K} \to \pi \pi}$. To this end, we insist that $|T^0_{1 1}|^2$ and $|T_{K \overline{K} \to K \overline{K}}|^2$ peak at the same value of $P_0 = M_R = M_{1 1}$, and have also the same widths; see Fig.~\ref{TijP0}.
This indicates that $|T^0_{1 1}|$ to be replaced by $|T^0_{1 1} (M_R)| = (\sqrt 2 g_{f_0 K \overline{K}})^2 / M_R \Gamma_R$, which gives the effective coupling constant value as $g_{f_0 K \overline{K}} = 2.808 \; $GeV$ $, for $\Lambda = 1.351 \; $GeV$ $, $M_R = M_{1 1} = 974 \; $MeV$ $, and $\Gamma_R = 43.19 \; $MeV$ $. The same procedure for $T^0_{2 1} (s) \to T_{K \overline{K} \to \pi \pi}$ gives
\begin{align}
g_{f_0 \pi \pi} g_{f_0 K \overline{K}} &= -\frac{1}{2 \sqrt 6} M_{2 1} \Gamma_{2 1} \left| T^0_{2 1}(M_{2 1}) \right| \nonumber \\
&= -1.849 \; \text{GeV}^2,
\end{align}
which gives $g_{f_0 \pi \pi} = -0.659 \; $GeV$ $. The negative effective coupling constant manifests itself in the second panel of Fig.~\ref{F2}, where the imaginary part of $T$ becomes negative. In Fig.~\ref{ImT}, we have compared the exact (ChPT), calculated imaginary part of $T^0_{2 1} (s)$ (solid, red curve) with its Breit--Wigner approximation (dashed, red curve).

\subsection{$\gamma \gamma \to m_1 m_2$ cross-sections}

\subsubsection{$\gamma \gamma \to \pi^0 \pi^0, \pi^0 \eta$ cross-sections}

\begin{figure*}

\begin{center}
\includegraphics[scale = 0.70]{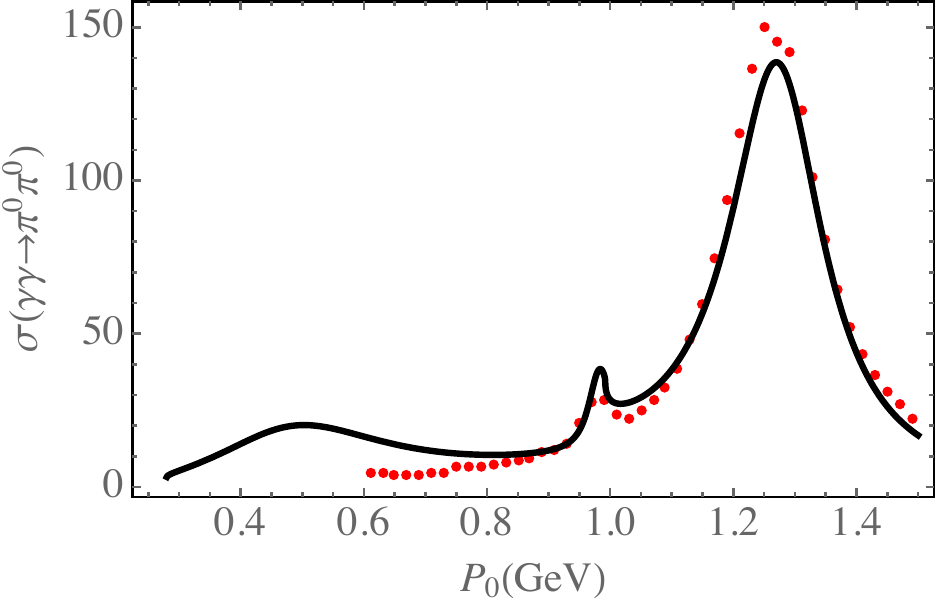}
\end{center}

\caption{Cross-section for $\gamma \gamma \to \pi^0 \pi^0$ integrated over the restricted angular range $|\cos\theta| < 0.6$ (solid, black curve) compared to the data of the Belle, Crystal Ball, and JADE Collaborations \cite{Be08, HM90, TOest90} (red dots).}

\label{F3}
\end{figure*}

The cross-section $\gamma \gamma \to \pi^0 \pi^0$ is often considered to be particularly instructive, as it lacks a Born term. Thus, the contributions from the $f_2$ and $a_2$ (as well as the possibilities that could arise from new physics) can be investigated more closely. As is explained in Sec.~\ref{sec2c}, due to the large total widths of these two resonances, they can be expected to contribute to production cross-sections well below their peak positions. In the $\gamma \gamma \to \pi^0 \pi^0$ channel the total cross-section given in (\ref{e:fullcross}) is just the sum of the partial cross-sections determined by ChPT and $f_2$ resonance amplitudes separately, without any interference term, as can be seen in Table~\ref{t:table2}. The results are shown in Fig.~\ref{F3}, where the calculated total cross-section, including the pure $(J, \lambda) = (0, 0)$ chiral contribution as well, are compared with the Belle \cite{Be08} as well as the older Crystal Ball and JADE Collaborations data \cite{HM90, TOest90}.
The transition amplitudes for these calculations come from (\ref{e:FSIhel}) and (\ref{e:resonanceamp}) for $T^{(0, 0)}_{S; \gamma \gamma \to \pi^0 \pi^0}$ and $T^{(2, 2)}_{f_2; \gamma \gamma \to \pi^0 \pi^0}$, evaluated at the cutoff $\Lambda = 1.351 \; $GeV$ $ in the first case, and using a mass, total width, and branching ratio of $(M_{f_2}, \Gamma_{f_2}) = (1275, 185) \; $MeV$ $ and $\Gamma^{(2)}_{\gamma \gamma} \text{Br} (\pi^0 \pi^0) \Gamma_{f_2} = 0.16 \; $MeV$^2 $ for the second resonance amplitude, as extracted from the PDG tables \cite{PDG10}. Note that if the experimental observations are restricted to $\theta_c < \theta < \pi - \theta_c$, the total cross-sections for a given angular momentum $J$ are modified by a factor $F_J(z)$, which reads $F_J(z) = \int_0^{2 \pi} d\phi \int_{\theta_c}^{\pi - \theta_c} d\theta \sin\theta |Y_{J, J}(\theta, \phi)|^2$.

One notes that the ChPT cross-section is lifted sufficiently in the vicinity of $\sim 1 \; $GeV$ $ by the low energy tail of the $f_2$ resonance contribution to lead to an acceptable overall fit with the experiment. This result, in turn, confirms that the chiral $(0, 0)$ cross-section is valid for center-of-mass energies below $\sim 1 \; $GeV$ $.

\begin{figure*}

\begin{center}
\includegraphics[scale = 0.70]{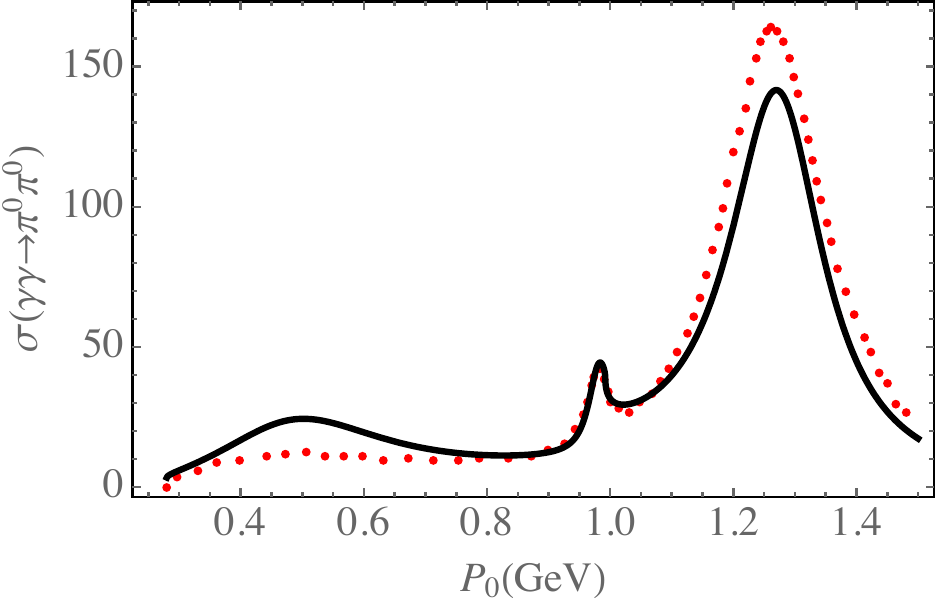}
\end{center}

\caption{Cross-section for $\gamma \gamma \to \pi^0 \pi^0$ integrated over the full angular range (solid, black curve) compared to the extracted curve, denoted in their paper as Solution I, of Dai and Pennington \cite{DaiPenn14} (red dots).}

\label{F4}
\end{figure*}

The cross-section for $\gamma \gamma \to \pi^0 \pi^0$ integrated over the full angular range is shown in Fig.~\ref{F4}. Here one sees again that the resonance $f_2$ underestimates the results of Dai and Pennington \cite{DaiPenn14} in its strength, and some discrepancy is also observed at lower energies.

\begin{figure*}

\begin{center}
\includegraphics[scale = 0.70]{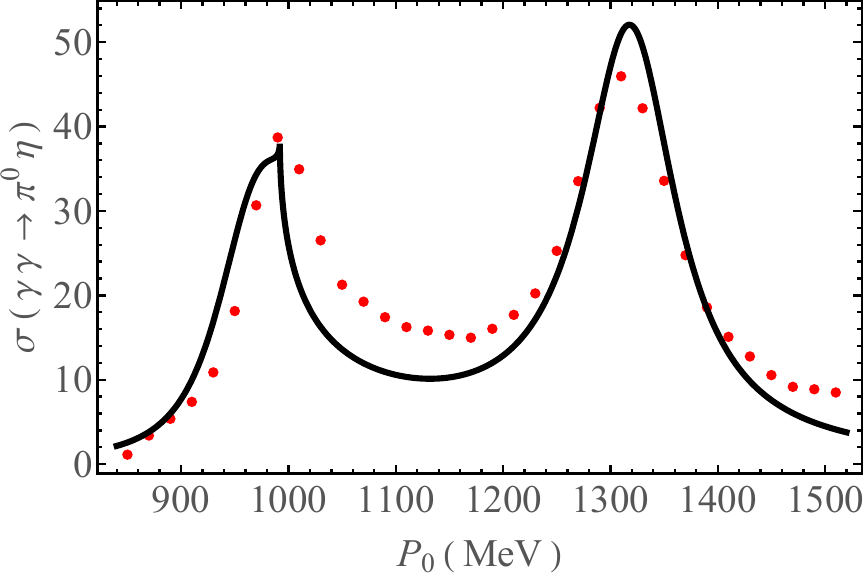}
\end{center}

\caption{Cross-section for $\gamma \gamma \to \pi^0 \eta$ integrated over the angular range $|\cos\theta| < 0.8$ (black curve) compared to the data of the Belle Collaboration \cite{Be09} (red dots).}

\label{F9}
\end{figure*}

In a similar fashion to the $\gamma \gamma \to \pi^0 \pi^0$, the cross-section for the process $\gamma \gamma \to \pi^0 \eta$ can be calculated; the result is shown in Fig.~\ref{F9}. Our theoretical calculation (black curve) can capture the essential structure of the Belle Collaboration data \cite{Be09} quite well, although not perfect.

\subsubsection{$\gamma \gamma \to \pi^+ \pi^-$ cross-section}

\begin{figure*}

\begin{center}
\includegraphics[scale = 0.70]{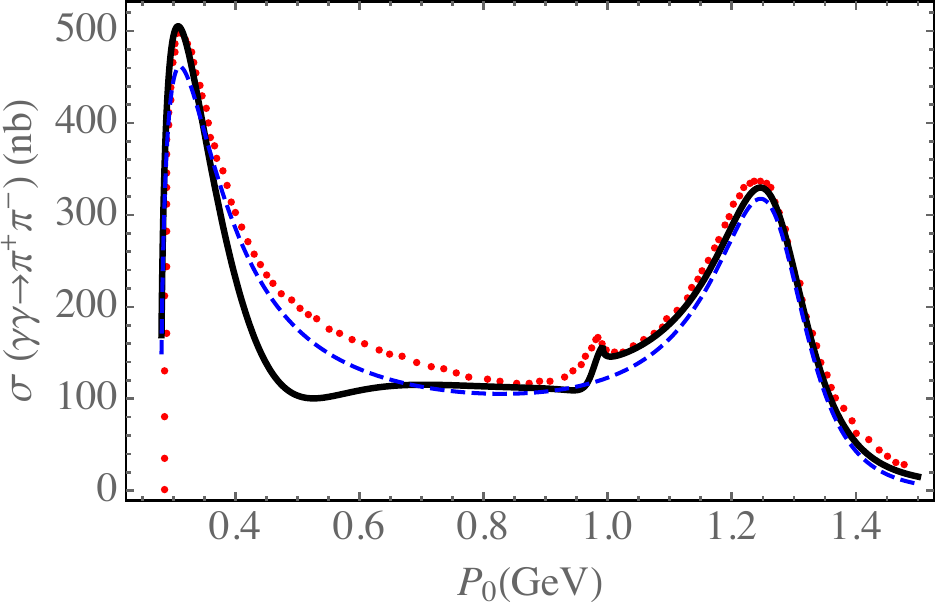}
\end{center}

\caption{Cross-section for $\gamma \gamma \to \pi^+ \pi^-$ integrated over the full angular range (solid, black curve) compared to the extracted curve, denoted in their paper as Solution I, of Dai and Pennington \cite{DaiPenn14} (red dots). The blue, dashed curve is a calculation not including the effects of the strong interaction.}

\label{F5}
\end{figure*}

In Fig.~\ref{F5}, we have shown the results for the process $\gamma \gamma \to \pi^+ \pi^-$.
In this figure, the dotted curve represents the extracted data of Dai and Pennington \cite{DaiPenn14}, integrated over the full angular range. The blue, dashed curve is our calculation, without including the effects of the final-state strong interactions, while the black, solid curve indicates our final full calculation, including the final-state strong interactions. As expected, the Born contribution to this process plays a dominant role at low energies. After 0.9 GeV, this calculation is in very good agreement with the extracted data.

\section{Discussion and outlook}
\label{sec4}

Calculations of the photon--photon to meson--meson cross-sections via various theoretical approaches at different levels have been addressed by many authors in the past. Just to mention a few examples: the authors of \cite{MH2011} construct a complete set of Roy--Steiner equations for the $\gamma \gamma \to \pi \pi$ reaction. Using the proposed formalism and approximating the $f_2 (1270)$ resonance by a Breit--Wigner ansatz, they calculate the cross-section of pion pair production for both charged and neutral pions. A chiral Lagrangian model, with dynamical light vector mesons, is presented in \cite{IVD} to study the production of $\pi \pi$, $\pi \eta$, and $K \overline{K}$ in photon--photon collisions and to evaluate the associated cross-sections. The proposed model of \cite{IVD} does not incorporate the $f_2 (1270)$ and $a_2 (1320)$ resonances. Using Muskhelishvili--Omn\`es (MO) dispersive representations of photon--photon scattering to two pions, the authors of \cite{RGM2010} evaluate the cross-section for pion pair production, including the contribution of tensor resonances. Their study, however, does not provide an analysis of the production of other meson pairs such as the $\pi \eta$. The $\gamma \gamma \to \pi \eta$ reaction is investigated in \cite{ID2017}, using the $S$-matrix theory. Although the authors consider the effect of the $a_2 (1320)$ resonance through a Breit--Wigner approximation, their calculation does not include the axial vector resonance exchange contribution. Within the realm of chiral perturbation theory (ChPT), the cross-section for $\gamma \gamma \to \pi \pi$ is evaluated at one-loop order in \cite{BiCo88}, where both charged pion and neutral pion pair production are studied. A similar approach for the $\pi^0 \pi^0$ production is presented in \cite{JFD1988}. The reported results of \cite{BiCo88, JFD1988} for $\gamma \gamma \to \pi^0 \pi^0$ show disagreement with the Crystal Ball data. Two-loop order calculation within ChPT for the $\pi^0 \pi^0$ production, as given in \cite{SB1994, JG2005}, improves agreement with the experiment. For the case of the $\pi^+ \pi^-$ production, the corresponding calculation at two-loop order is given in \cite{UB1996, JG2006}. Note that at this order, most of the coupling constants of the chiral Lagrangian are still undetermined \cite{RGM2010}. Also, the studies of \cite{BiCo88, JFD1988, JG2005, UB1996, JG2006} do not take into account the $f_2 (1270)$ resonance, and its effects have been ignored. All the papers mentioned above concern themselves mainly with reproducing experimental data for cross-sections; however, they lack a detailed amplitude analysis. In this paper, we have recalculated these cross-sections, using a version of ChPT, with two aims in mind: firstly, we wish to make a comparison at the level of the transition amplitudes and not only directly for the cross-sections. Secondly, we have in mind to investigate in the future the possibility of the formation and decay of kaonic atoms. Should these exist, they would be expected to have an extremely short lifetime, but a large cross-section. For this, we need to have a reasonable agreement for the cross-sections over a range somewhat above 1 GeV and the use of ChPT is essential. We have thus calculated the amplitudes and cross-sections for photon--photon collisions giving rise to $\pi^+ \pi^-$, $\pi^0 \pi^0$, $\pi^0 \eta$, and $K \overline{K}$ in the final state. Our theoretical framework combines the Born scattering transition amplitudes required through QED with those that can be calculated via ChPT to describe the low-energy regime and includes parametrizations of the resonant mesons such as $f_2(1270)$ and $a_2(1320)$, which cannot be accounted for within ChPT, but which are essential for the evaluation of the cross-sections. We have compared our results with those from \cite{DaiPenn14}, which are extracted from the high statistics data from Belle, as well as the older data from Mark II at SLAC, CELLO at DESY, and Crystal Ball at SLAC, and fitted using basic constraints of analyticity, unitarity, and crossing symmetry, as well as Low's low-energy theorem for QED. It is noteworthy that the authors of \cite{DaiPenn14} are not only able to fit the cross-sections but also able to obtain the strong-interaction transition matrices, to which we have been able to compare our theoretical model. The results of the comparison are reasonable, showing all expected structures, but they are not perfect. It is an open question as to whether the low-energy behavior can be improved substantially if the experimentally inferred inputs for ChPT, such as the combination $L_9 + L_{10}$ are better known, or if some new effects, for example, due to mesonic substructure are evident. We also note that in the extracted amplitudes in \cite{DaiPenn14}, denoted as Solution I, the $f_0 (1370)$ appears at the very edge of where the analysis of \cite{DaiPenn14} with just $\pi \pi$ and $K \overline{K}$ channels can be trusted. Thus, we have not incorporated this state into our approach.

For our intentions, the cross-sections calculated in this fashion are sufficient to provide a basis for addressing further intriguing questions, such as whether other new structures like the postulated existence of the kaonium atom $K^+ K^-$ actually exist and can be observed. This can be addressed within the context of the current paper, for which these cross-sections are required. In a nutshell, we will look for the kaonium as a sharp resonance possibly accompanying the $f_0$ in the process $\gamma \gamma \to \pi^0 \pi^0$, or in the $\gamma \gamma \to \pi^0 \eta$. For example, the cross-section for $\gamma \gamma \to \pi^0 \pi^0$, including the formation of kaonium, can be written as
\begin{align}
&\sigma(\gamma \gamma \to \pi^0 \pi^0) = \frac{1}{256 \pi s} \sqrt{1 - \frac{4 m_{\pi}^2}{s}} \left( \frac{\alpha}{\pi} \right)^2 \nonumber \\
&\times \left| J_{\pi}(s) T_{\pi^+ \pi^-; \pi^0 \pi^0}(s) + J_K(s) T_{K^+ K^-; \pi^0 \pi^0}(s) \right|^2,
\end{align}
where $J_{\pi}(s)$ and $J_K(s)$ can be calculated from (\ref{e:JpiK}). The charge exchange $T$-matrix for $s$-wave pions is found to be $T_{\pi^+ \pi^-; \pi^0 \pi^0}(s) = \langle \pi^0 \pi^0 | T | \pi^+ \pi^- \rangle = ( T^0_{2 2}(s) - T^2_{2 2}(s) ) / 3$, and the annihilation amplitude $\langle \pi^0 \pi^0 | T | K^+ K^- \rangle$ into two neutral pions reads $T_{K^+ K^-; \pi^0 \pi^0}(s) = \langle \pi^0 \pi^0 | T | K^+ K^- \rangle = 1 / \sqrt 6 T^0_{2 1}(s)$.

However, the subtle point is that due to the $K^+ K^-$ Coulomb interaction, which is required for kaonium formation, there is an isospin breaking which should be imported in $T_{K^+ K^-; \pi^0 \pi^0}(s)$ ($T_{\pi^+ \pi^-; \pi^0 \pi^0}(s)$ is not affected by the isospin breaking in the $K^+ K^-$ channel). In short, we will import the effects of isospin breaking into the transition amplitudes via the value of the modified strong inverse scattering length of $K^+ K^-$ in the presence of attractive Coulomb fields. Considering this effect, the cross-section for $\gamma \gamma \to \pi^0 \pi^0$ and also for $\gamma \gamma \to \pi^0 \eta$, including the formation of kaonium, can be evaluated. One would expect the kaonium resonance to manifest itself as a sharp peak around 980 MeV. The detailed discussion and calculation of these will be reported in our future paper.

\end{document}